\documentclass{IEEEjmw}

\usepackage{color,array}
\usepackage{xcolor}

\usepackage{amsmath,amssymb,amsfonts}
\usepackage{siunitx}

\usepackage{graphicx}
\usepackage{tikz}
\usepackage[american]{circuitikz}
\usepackage[caption=false,font=footnotesize]{subfig}

\usepackage{booktabs}
\usepackage{makecell}
\usepackage{float}

\usepackage{algorithmic}
\usepackage[colorlinks,urlcolor=blue,linkcolor=blue,citecolor=blue]{hyperref}

\makeatletter

\def\ps@headings{%
  \def\@oddhead{}%
  \def\@evenhead{}%
  \def\@oddfoot{\hfil\thepage\hfil}%
  \def\@evenfoot{\hfil\thepage\hfil}%
}

\def\ps@IEEEtitlepagestyle{%
  \def\@oddhead{}%
  \def\@evenhead{}%
  \def\@oddfoot{\hfil\thepage\hfil}%
  \def\@evenfoot{\hfil\thepage\hfil}%
}

\def\ps@titlepagestyle{%
  \def\@oddhead{}%
  \def\@evenhead{}%
  \def\@oddfoot{\hfil\thepage\hfil}%
  \def\@evenfoot{\hfil\thepage\hfil}%
}

\makeatother

\makeatletter
\def\@history{}
\def\@receiveddate{}
\def\@reviseddate{}
\def\@accepteddate{}
\def\@publisheddate{}
\def\@currentdate{}
\makeatother

\makeatletter

\let\old@maketitle\maketitle
\renewcommand{\maketitle}{%
  \begingroup
  \def\@receiveddate{}%
  \def\@reviseddate{}%
  \def\@accepteddate{}%
  \def\@publisheddate{}%
  \def\@currentdate{}%
  \def\@history{}%
  \def\history##1{}%
  \old@maketitle
  \endgroup
}

\makeatother
\begin{document}

\title{A Microfabricated PCM-Switched Reconfigurable Intelligent Surface for Wideband Millimeter-Wave Beam Steering}


\author{Afsaneh~Hojjati-Firoozabadi\affilmark{1} (Graduate Student Member, IEEE)}
\author{Raafat~Mansour\affilmark{1} (Life Fellow, IEEE)}

\affil{\affilmark{1}Department of Electrical and Computer Engineering, University of Waterloo, Waterloo, ON N2L~3G1, Canada}

\corresp{CORRESPONDING AUTHOR: A. Hojjati-Firoozabadi (e-mail: \href{mailto:a4hojjat@uwaterloo.ca}{a4hojjat@uwaterloo.ca}).}

\authornote{This work was conducted at the Centre for Integrated RF Engineering (CIRFE), University of Waterloo. 
\\{\normalsize\bfseries This work has been submitted to the IEEE for possible publication. Copyright may be transferred without notice, after which this version may no longer be accessible.}}


\begin{abstract}
This paper presents the design, fabrication, and experimental validation of a reconfigurable intelligent surface (RIS) employing electrically actuated vanadium dioxide (VO$_2$) switches for millimeter-wave beam steering. The proposed RIS is realized using a multilayer microfabrication process on an alumina substrate, enabling monolithic integration of hundreds of sub-4~$\mu$m VO$_2$ switching elements within deeply subwavelength unit cells ($\sim\lambda/5.2$). The switching-induced modulation of surface impedance is analyzed through full-wave simulations, and the resulting phase and amplitude responses are experimentally characterized using a custom WR-28 waveguide measurement setup. Based on the validated unit-cell design, a $10 \times 20$ RIS array integrating 200 VO$_2$ switches is fabricated. The switches within each column are serially biased using integrated routing lines, allowing programmable control of the spatial phase distribution across the surface. Synthesized phase profiles enable dynamic beam steering, resulting in measured far-field gain enhancement of 10--20~dB over an 18\% fractional bandwidth centered at 33~GHz, with steering angles up to $60^\circ$. The measured radiation patterns are in good agreement with semi-numerical channel modeling predictions. 
By combining thin-film PCM switching with an integration-aware unit-cell design, this work demonstrates a scalable RIS architecture that mitigates packaging parasitics and footprint limitations inherent to conventional semiconductor-based implementations, providing a practical pathway toward higher-frequency reconfigurable surfaces.
\end{abstract}


\maketitle

\begin{tikzpicture}[remember picture,overlay]
\fill[white] ([xshift=-3cm,yshift=0.5cm]current page.north) rectangle
             ([xshift=7cm,yshift=0.0cm]current page.north);
\end{tikzpicture}

\makeatletter
\def\@oddhead{}%
\def\@evenhead{}%
\def\@oddfoot{\hfil\thepage\hfil}%
\def\@evenfoot{\hfil\thepage\hfil}%
\makeatother
\thispagestyle{empty}

\vspace{-0.7cm}
\section{INTRODUCTION}
Reconfigurable intelligent surfaces (RISs) have emerged as a powerful paradigm for electromagnetic (EM) wave engineering, enabling programmable control over the phase, amplitude, and polarization of incident waves. By shaping wavefronts through spatially distributed, electronically reconfigurable unit cells, RISs can enhance signal coverage, enable beam steering, and mitigate interference in complex propagation environments. These capabilities are particularly relevant for 5G and beyond-5G wireless systems, where operation at millimeter-wave (mmWave) and sub-terahertz frequencies suffers from severe path loss, blockage, and sensitivity to environmental dynamics. Beyond wireless communications, RISs have also been explored for wireless power transfer, sensing, localization, and energy-efficient smart radio environments. Among the wide range of wave manipulation functionalities enabled by RISs, far-field gain enhancement and beam steering are two fundamental capabilities that underpin many of these emerging applications~\cite{b1,b3,ris2025}.

The choice of the tunable element becomes increasingly critical for RIS implementations operating at millimeter-wave and beyond-millimeter-wave frequencies. In this regime, deeply subwavelength unit cells are required to enable dense phase sampling and precise beam angle control, causing integration constraints to dominate system performance~\cite{sub1}. Semiconductor PIN and varactor diodes have been widely employed in RIS and reflectarray designs at microwave and millimeter-wave bands~\cite{a1,a2,cite1,cite2,cite3,cite13}. However, their suitability for densely sampled RIS architectures becomes increasingly limited as operating frequency increases, due to a combination of physical size, parasitic effects, and hybrid integration requirements ~\cite{cite4}. In particular,
\begin{itemize}
    \item \textit{Physical size and scaling:} Typical packaged or effectively integrated PIN and varactor diodes occupy footprints on the order of several hundred micrometers, which do not scale with wavelength and become comparable to deeply subwavelength unit-cell dimensions at millimeter-wave frequencies (e.g., $\lambda/5 \approx 2$~mm at 30~GHz). In addition, practical implementations often require auxiliary structures such as RF chokes, vias, and clearance regions for bias routing, which further consume valuable unit-cell area and constrain the achievable geometry and miniaturization of the unit cell~\cite{cite2,cite3}. 
    \item \textit{Parasitic effects introduced by hybrid integration:} Hybrid integration of discrete semiconductor switches typically relies on soldering, wire bonding, or similar interconnect techniques. At millimeter-wave frequencies, these interconnects introduce parasitic inductance and capacitance that degrade phase accuracy, reduce operational bandwidth, and contribute additional insertion loss, with their impact becoming increasingly pronounced as frequency increases ~\cite{cite1,cite3}. \item \textit{Integration density and scalability:} Densely sampled RIS architectures require precise placement and alignment of a large number of discrete components within tightly constrained unit-cell footprints. As unit-cell dimensions shrink and array size increases, achieving the required alignment accuracy, repeatability, and fabrication yield across large apertures becomes challenging, imposing scalability limitations on hybrid-assembled RIS implementations.
\end{itemize}

Semiconductor RF switches based on GaAs, CMOS, and GaN technologies offer excellent intrinsic high-frequency performance and have therefore been explored as tuning elements for reconfigurable reflectarrays~\cite{cite5, cite6,cite7}. However, implementations relying on advanced semiconductor platforms face challenges in terms of fabrication complexity, cost, and scalability when extended to large apertures. These approaches typically require compound-semiconductor epitaxy or deep-submicron lithography, along with wafer-size and reticle limitations that complicate dense tiling and large-area integration. As a result, while semiconductor-based switches are well suited for compact or chip-scale implementations, their extension to large, densely sampled RIS architectures remains economically and technologically challenging at millimeter-wave and sub-terahertz frequencies.

RF microelectromechanical systems (MEMS) switches have also been explored for reconfigurable reflectarrays and RIS platforms due to their low insertion loss and high linearity~\cite{b6,mems1,mems3}. Despite these advantages, their practical deployment in densely integrated RIS architectures is constrained by integration-related factors. RF MEMS switches rely on mechanically actuated structures that require multilayer fabrication, sacrificial release processes, and suspended elements, which increase fabrication complexity and impact yield. In addition, bias routing, actuation electrodes, and mechanical anchors impose layout constraints that limit achievable unit-cell miniaturization. As a result, most reported MEMS-based reflectarrays operate with unit-cell dimensions on the order of $\lambda/2$ (and occasionally $\lambda/3$), predominantly at microwave frequencies~\cite{b_2,mems2}. The combination of mechanical reliability considerations, packaging requirements, and limited scaling capability has so far hindered the realization of densely sampled, large-scale RIS arrays based on RF MEMS technology at mmWave frequencies.

Phase-change materials (PCMs) have recently attracted significant attention as an alternative switching platform for RIS applications, owing to their favorable RF properties at millimeter-wave and sub-terahertz frequencies and their compatibility with standard microfabrication processes. PCMs enable direct monolithic integration of highly miniaturized tunable elements within the RIS unit cell, thereby minimizing interconnect parasitics, avoiding hybrid assembly steps, and facilitating deeply subwavelength unit-cell dimensions at mmWave frequencies.

Non-volatile and low-loss PCM switches based on materials such as GeTe and GST have been explored for reconfigurable RF and electromagnetic applications. While these PCM switches offer memory functionality, review studies report that their reliance on high-temperature phase transitions and precisely controlled programming pulses can pose challenges related to thermal management and large-area uniformity when scaling toward densely integrated RIS implementations~\cite{pcmris}.

Vanadium dioxide (VO$_2$), as another well-known PCM, offers a complementary and scalable approach for reconfigurable switching at mmWave frequencies and beyond. VO$_2$ exhibits a sharp and reversible metal--insulator transition (MIT), with a conductivity change exceeding three orders of magnitude~\cite{a44}, and can be actuated through thermal, optical, or electrical mechanisms~\cite{a66}. While thermally driven transitions require integrated microheaters and careful thermal design~\cite{a77}, electrically driven MIT (E-MIT) enables simpler device structures with reduced process complexity and improved scalability, making it particularly attractive for RIS integration.

Importantly, VO$_2$ switches can be realized with lateral dimensions of only a few micrometers and patterned directly within the RIS unit cell, enabling high integration density without the packaging parasitics associated with hybrid integrations. Standalone microfabricated VO$_2$ switches have demonstrated low insertion loss, high isolation, and linear behavior at mmWave frequencies extending beyond 60~GHz~\cite{aa77}, with reported $R_\text{ON}C_\text{OFF}$ figures of merit in the multi-terahertz range.

Despite the demonstrated potential of VO$_2$-based switches, only a limited number of studies have reported experimentally implemented VO$_2$-based reconfigurable surfaces for beam steering. In contrast, the majority of demonstrated RIS architectures to date rely on semiconductor diode-based switching elements. 

Early work demonstrated a thermally tuned transmissive metasurface that achieved a limited continuous phase shift by modulating the permittivity of VO$_2$ using integrated microheaters~\cite{a70}. However, this approach did not exploit binary conductivity switching, operated VO$_2$ within its transition region—resulting in high loss—and was limited to a small array size and a narrow steering range.

A more recent reflective RIS implementation employed screen-printed VO$_2$ operating at 23.5--29.5\,GHz~\cite{a71}. While cost-effective, screen printing inherently produces coarse feature sizes and lower-quality VO$_2$ films, leading to oversized and lossy switching regions that limit scalability toward higher frequencies. A subsequent work extended screen-printed VO$_2$ switches to higher-frequency operation in a large RIS configuration~\cite{vo3}; however, the unit-cell dimensions remained on the order of $\lambda/2$, preventing subwavelength sampling and effectively placing the surface operation closer to a reflectarray regime rather than a densely sampled RIS. Although these studies represent important steps toward VO$_2$-based reconfigurable surfaces, they do not fully exploit the potential of PCM-enabled binary switching nor the advantages offered by monolithic microfabrication for densely sampled beam-steering RIS implementations.

In another work, a microfabricated VO$_2$-based reflective surface employing complementary unit-cell designs demonstrated beam switching~\cite{vo4}. However, this surface was actuated through global thermal control, which precludes independent phase programmability across the aperture and therefore lacks the reconfigurability required for continuous beam steering. Moreover, despite the use of microfabrication, the VO$_2$ switching elements were not miniaturized and exhibited feature sizes comparable to those obtained via screen-printing techniques, limiting achievable integration density and scalability. This outcome underscores the importance of lithographic precision and material quality not only for low-loss high-frequency operation, but also for enabling dense integration and the electrically programmable phase control required for beam-steering RIS architectures.

Nevertheless, no prior work has demonstrated a fully monolithic, microfabrication-enabled RIS that integrates hundreds of micron-scale PCM switches within deeply subwavelength unit cells while enabling electrically programmable, spatially varying phase control for beam steering.
 Motivated by these limitations, this work addresses the integration-limited nature of RIS scalability at millimeter-wave frequencies, where switch footprint, parasitic effects, and fabrication scalability become dominant constraints.

 Building on our earlier studies~\cite{a72,a73}, we demonstrate a fully microfabricated RIS architecture enabled by monolithically integrated, miniaturized VO$_2$ switches. The proposed RIS integrates 200 sub-4~$\mu$m VO$_2$ switches into deeply subwavelength unit cells ($\lambda/5.2$), achieves wideband binary phase control validated through WR28 waveguide measurements, and experimentally demonstrates beam steering up to $\pm60^\circ$ with more than 10~dB far-field gain enhancement over an 18\% fractional bandwidth centered at 33~GHz. By treating the switching element as a lithographically defined component of the unit-cell topology rather than a hybrid add-on, this work establishes the hardware feasibility and scalability of a monolithic PCM-based RIS architecture suitable for integration-limited, high-frequency, densely sampled RIS architectures.

The remainder of the paper is organized as follows. Section~\ref{sec:unit_cell_design} presents the unit-cell design and its surface-impedance-based analysis. Section~\ref{sec:supercell_waveguide} reports waveguide characterization of a supercell with integrated VO$_2$ switches and introduces the biasing and protection circuitry. Section~\ref{sec:ris_design_simulation} describes the design and numerical analysis of the $10\times20$ RIS array. Section~\ref{sec:ris_fabrication_measurement} details the microfabrication process and measurement setup, while Section~\ref{sec:measurement_results} presents the far-field beam-steering measurement results. Section~\ref{sec:discussion} discusses and compares the results with prior works, and Section~\ref{sec:conclusion} concludes the paper.

\section{Unit Cell Design}
\label{sec:unit_cell_design}

\begin{figure}[t]
\centering
\vspace{-0.1cm}

\subfloat[]{
  \centering
  \includegraphics[width=80mm]{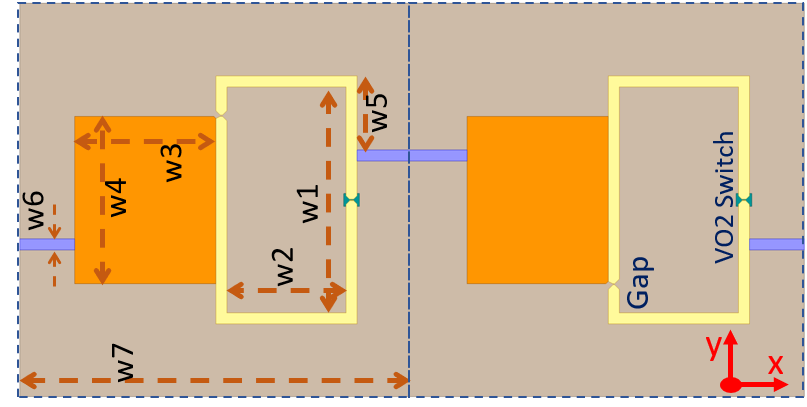} 
  \label{fig:unitcell1-a}
}

\vspace{0.2cm} 

\subfloat[]{
  \centering
  \includegraphics[width=70mm]{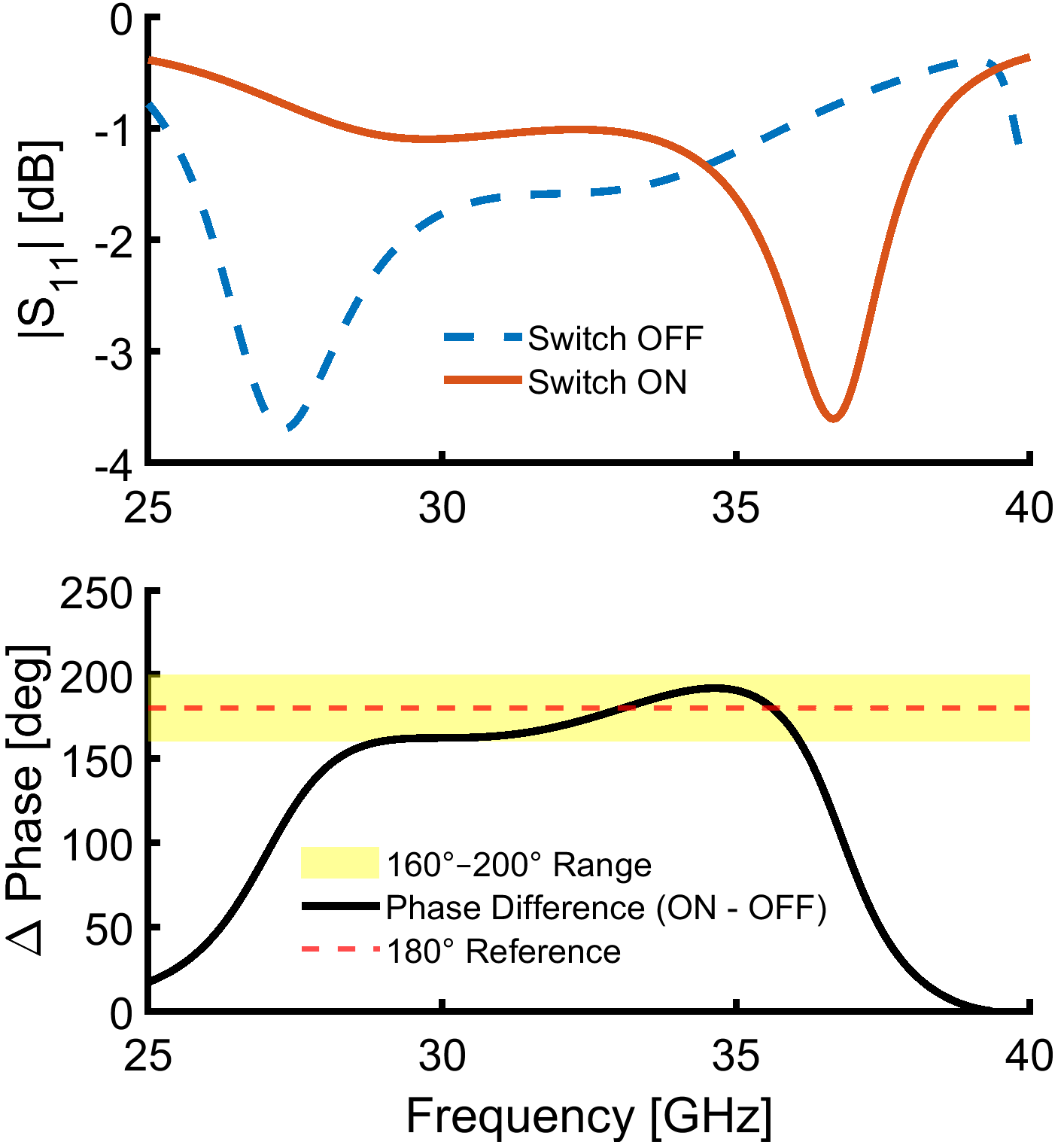} 
  \label{fig:unitcell1-b}
}

\caption{(a) Unit-cell layout with labeled parameters. 
(b) Simulated $|S_{11}|$ and phase difference for ON and OFF states under normal incidence. }
\label{fig:unitcell1}
\vspace{0.02cm}
\end{figure}

A deeper understanding of the structure introduced in our previous work~\cite{a73} is provided here. As illustrated in Fig.~\ref{fig:unitcell1}\subref{fig:unitcell1-a}, the unit cell consists of three main components: an asymmetric split-ring resonator (SRR) with two gaps—one of which is replaced by a \textit{VO}$_2$ switch—a metallic patch, and biasing lines, each highlighted in a different color in the figure. The geometric dimensions of the unit cell components are listed in Table~\ref{tab:unitcell_params}. The entire structure was fabricated on an alumina substrate with a thickness of \SI{625}{\micro\meter}, and the backside of the substrate was fully metallized, enabling the unit cell to function as a reflective surface.

A closer inspection of the structure reveals that the unit cell is composed of two smaller subwavelength cells. The asymmetry in these smaller cells—particularly within the SRR—leads to undesired cross-polarized reflections when illuminated by a linearly $y$-polarized plane wave. To suppress this effect, two such sub-cells are placed adjacent to each other along the $x$-direction, forming a larger composite unit cell, as shown in Fig.~\ref{fig:unitcell1}\subref{fig:unitcell1-a}. One of the sub-cells is mirrored with respect to the other, resulting in the cancellation of cross-polarized components and thereby preserving the polarization of the reflected wave. As a result, the unit cell is designed to operate under $y$-polarized excitation.

This composite unit cell, formed by two mirrored sub-cells, exhibits two distinct states depending on whether the \textit{VO}$_2$ switch is in the ON or OFF state. The phase difference between these two states, along with the corresponding reflection magnitude, is shown in Fig.~\ref{fig:unitcell1}\subref{fig:unitcell1-b}
, under normal incidence of a plane wave and periodic boundary conditions.

\begin{table}[b]
\centering
\renewcommand{\arraystretch}{1.3}
\setlength{\tabcolsep}{6pt}

\caption{Geometric parameters of the unit cell.}
\label{tab:unitcell_params}

\begin{tabular}{|>{\centering\arraybackslash}m{1.35cm}
                |>{\centering\arraybackslash}m{0.85cm}
                |>{\centering\arraybackslash}m{0.85cm}
                |>{\centering\arraybackslash}m{0.85cm}
                |>{\centering\arraybackslash}m{0.85cm}
                |>{\centering\arraybackslash}m{0.85cm}|}
\hline
\textbf{Parameter} & w1 & w2 & w3 & w4 & w5 \\
\hline
\textbf{Value (µm)} & 1088.5 & 622 & 622 & 734.8 & 325.6 \\
\hline
\textbf{Parameter} & w6 & w7 & w8 & Gap Size& Switch Size\\
\hline
\textbf{Value (µm)} & 48.6 & 1728 & 1728 & 4 & 4 \\
\hline
\end{tabular}
\end{table}

To understand the origin of the phase shift between the ON and OFF states, it is essential to examine the resonant behavior of the unit cell. Although the reflection coefficient $|S_{11}|$ provides general insight, the high reflectivity of the resonances often masks their presence in the magnitude response. To address this, the surface impedance $Z_s$ of the structure is extracted from the simulated reflection coefficient using the following relation:

\begin{equation}
    Z_s = \eta_0 \cdot \frac{1 + S_{11} e^{2 j \beta d}}{1 - S_{11} e^{2 j \beta d}}
\end{equation}

\noindent
where $\eta_0$ is the free-space intrinsic impedance, $d$ is the distance from the surface of the unit cell to the reference plane, and $\beta = 2\pi f / c$ is the propagation constant in free space. The exponential term applies a phase correction to $S_{11}$ to accurately represent the surface impedance at the unit-cell interface.

\begin{figure}[t]
\centering
\vspace{-0.5cm}
\subfloat[]{
\centering
\includegraphics[width=23mm]{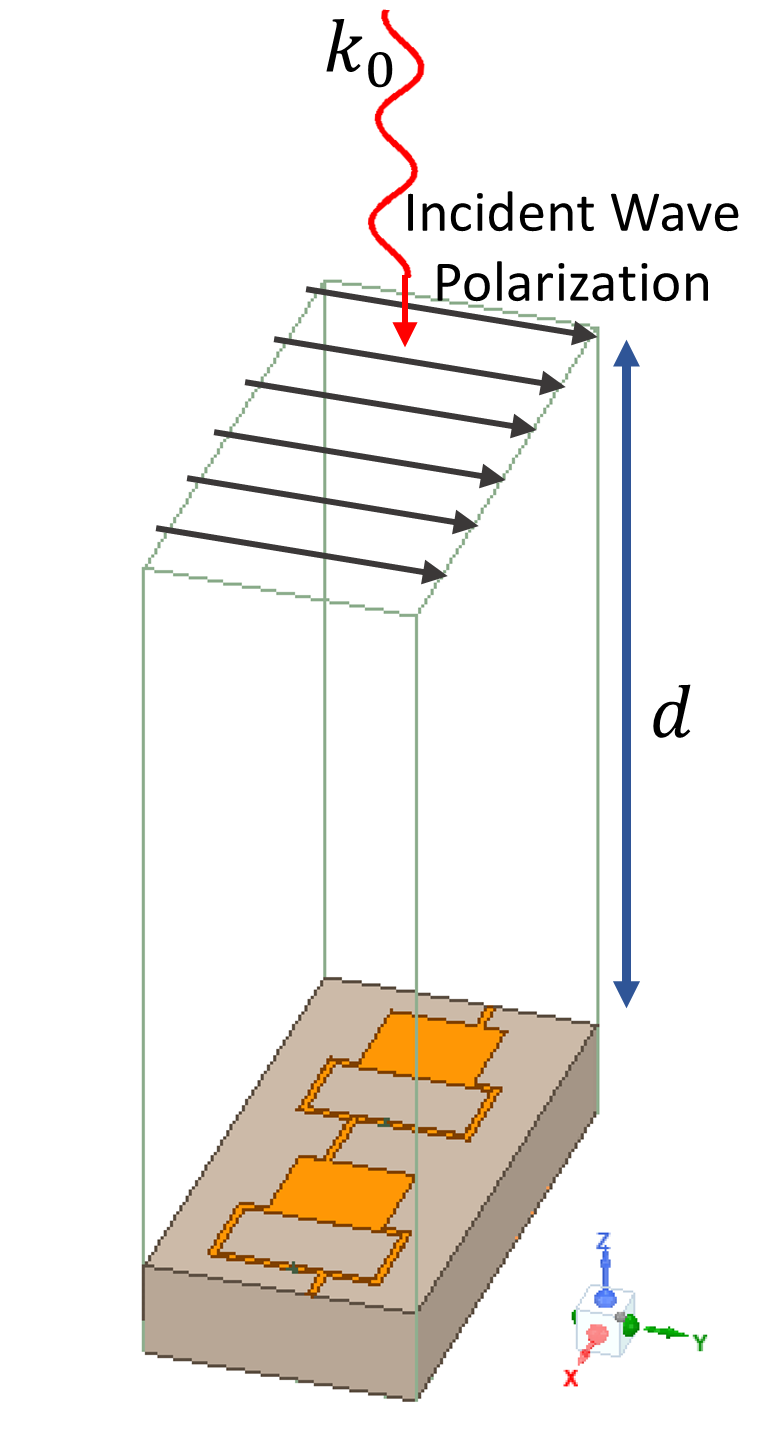} 
\label{fig:unitcell2-a}
}
~
\hspace{-0.5cm}
\subfloat[]{
\centering
\includegraphics[width=65mm]{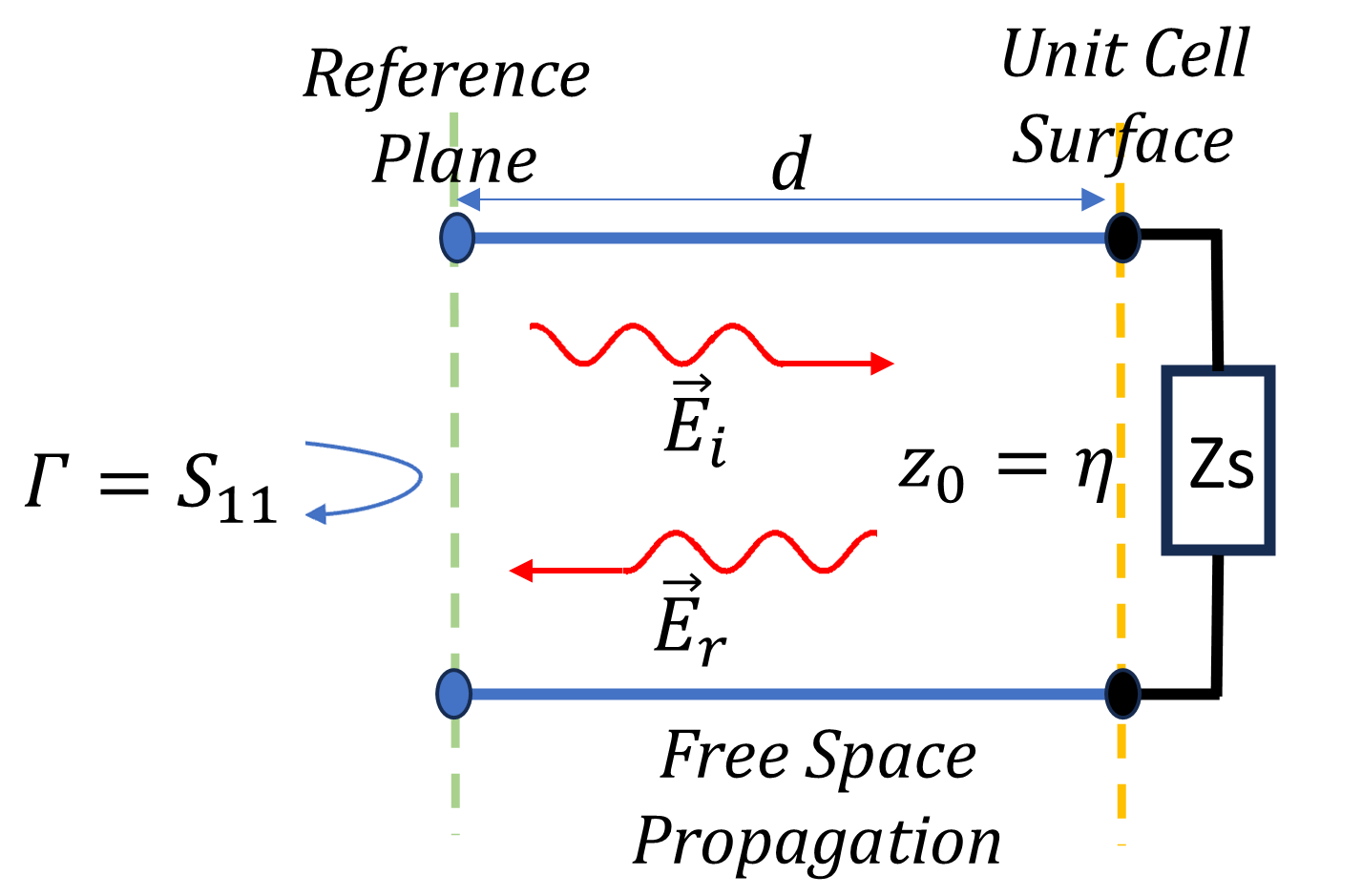} 
\label{fig:unitcell2-b}
}

\caption{(a) Unit cell simulated with periodic boundary conditions under normal plane-wave incidence. (b) Equivalent transmission-line circuit model.}

\vspace{0.02cm}
\label{fig:unitcell2}
\end{figure}

This formulation can be interpreted using transmission-line theory, where free space is modeled as a transmission line with characteristic impedance $\eta_0$, and the metasurface acts as a terminating impedance $Z_s$. In this analogy, $S_{11}$ corresponds to the reflection coefficient observed at the reference plane. The extracted surface impedance represents the ratio of tangential electric to magnetic fields at the interface, providing a more comprehensive understanding of the electromagnetic behavior of the unit cell than $S$-parameters alone. Fig.~\ref{fig:unitcell2}\subref{fig:unitcell2-a}
 and Fig.~\ref{fig:unitcell2}\subref{fig:unitcell2-b}
 illustrate the simulation setup and the corresponding transmission-line equivalent circuit, respectively.

\begin{figure}[t]
\centering

\subfloat[]{%
    \includegraphics[width=1\linewidth]{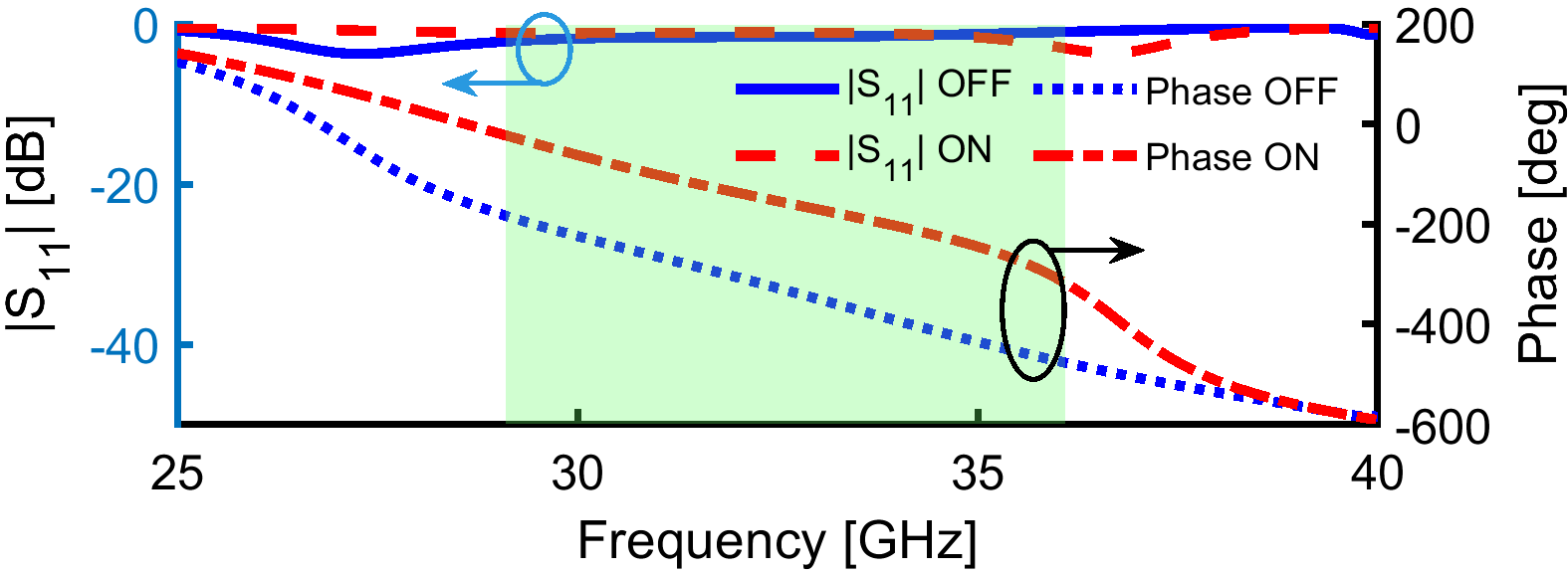}%
    \label{fig:a}%
}

\vspace{-3mm} 

\subfloat[]{%
    \includegraphics[width=1\linewidth]{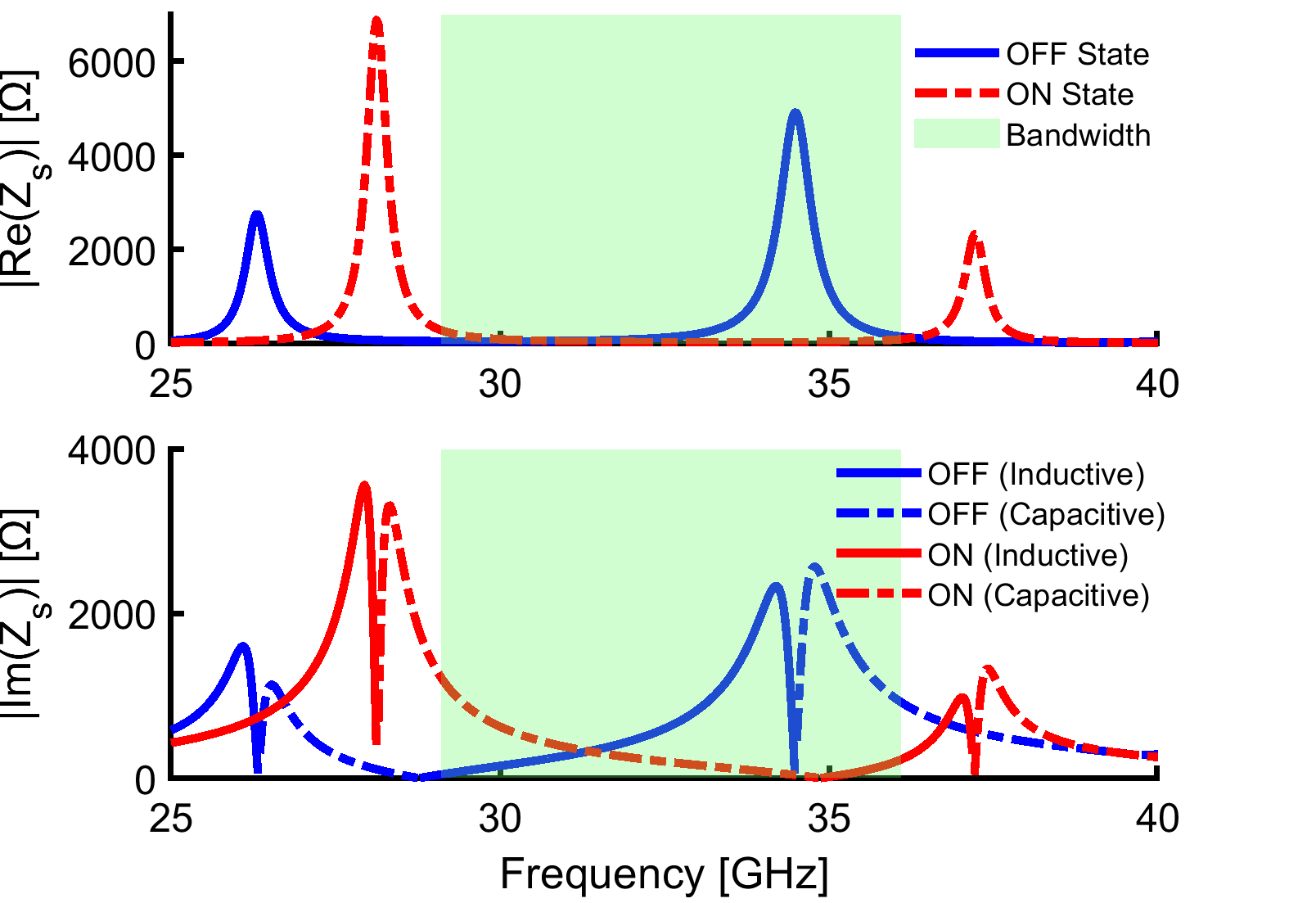}%
    \label{fig:b}%
}

\caption{(a) Simulated reflection magnitude and phase for ON and OFF states. (b) Real and imaginary parts of the extracted surface impedance for both states.}
\vspace{-0.3cm}
\label{fig:two_subfigures}
\end{figure}

Fig.~\ref{fig:two_subfigures} illustrates the reflection characteristics and extracted surface impedance of the proposed unit cell under normal plane-wave excitation. Fig.~\ref{fig:two_subfigures}\subref{fig:a}
 presents the reflection coefficient magnitude and phase for both the ON and OFF switching states. As observed, the structure maintains high reflectivity across the 25--40~GHz range while achieving a notable phase contrast between the two states—approximately $160^\circ$ to $195^\circ$ within the frequency band of 29.1--36.1~GHz. This corresponds to a 21.5\% fractional bandwidth centered around 32.6~GHz for the unit-cell response.

The real and imaginary components of the extracted surface impedance, shown in Fig.~\ref{fig:two_subfigures}\subref{fig:b}, reveal two distinct resonances for each switching state.
These resonances are characterized by peaks in the real part of $Z_s$ and corresponding zero crossings in the imaginary part. The observed phase difference between the two switching states is a direct consequence of these resonances. Although each state exhibits two resonances, one is more strongly excited—corresponding to a higher loaded quality factor—and produces a sharper phase transition in the reflection response, while the other, weaker resonance yields a smoother phase variation. Considering both states, the two stronger resonances—located near the edges of the operational bandwidth—primarily generate the required phase shift between the ON and OFF conditions, while the two weaker resonances help to tailor and maintain the desired phase difference across the frequency band.

\begin{figure}[b]
\centering

\subfloat[]{
\centering
\includegraphics[width=24mm]{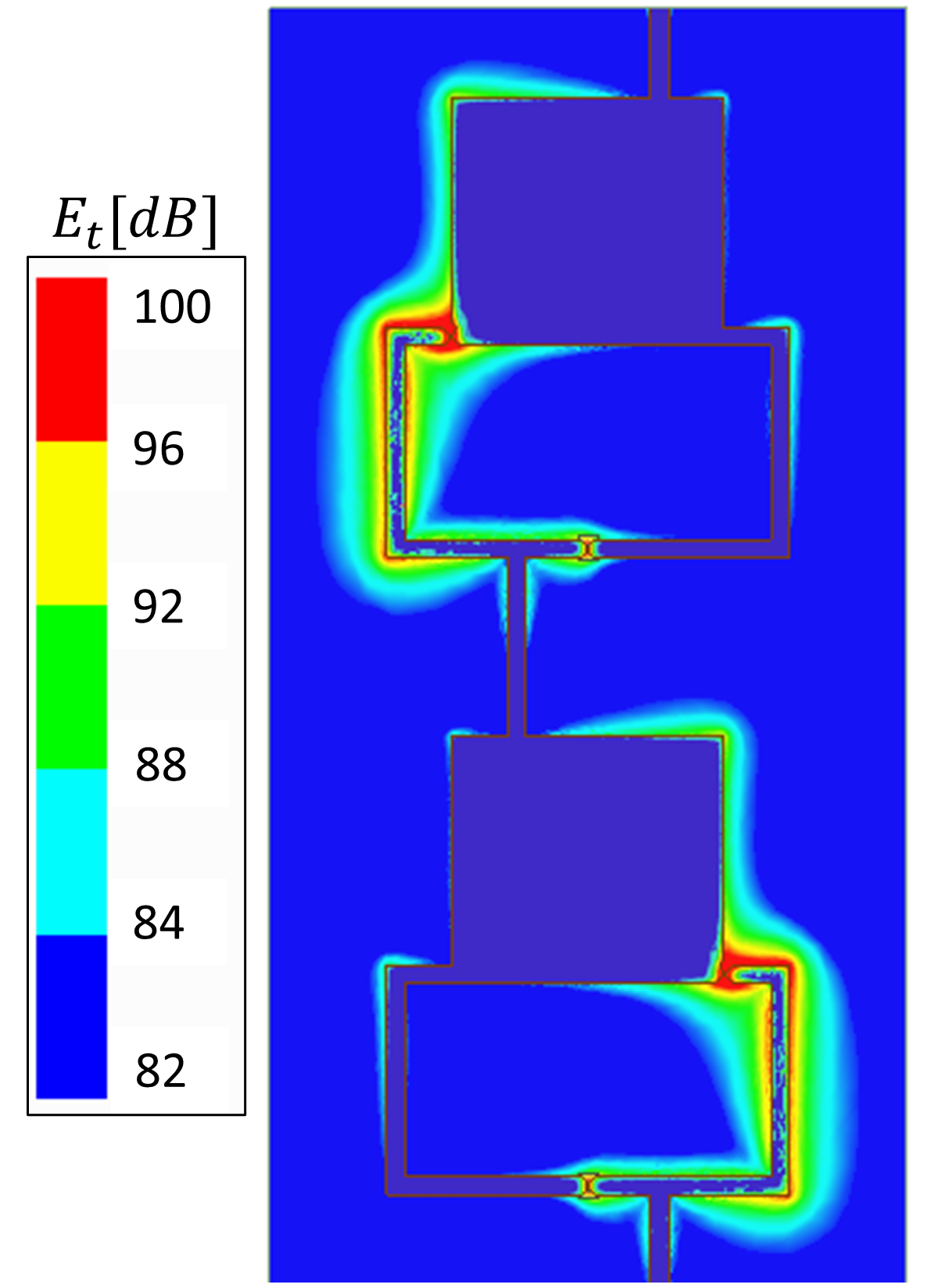} 
\label{fig:unitcell-a1}
}
~
\hspace{-0.5cm}
\subfloat[]{
\centering
\includegraphics[width=16.5mm]{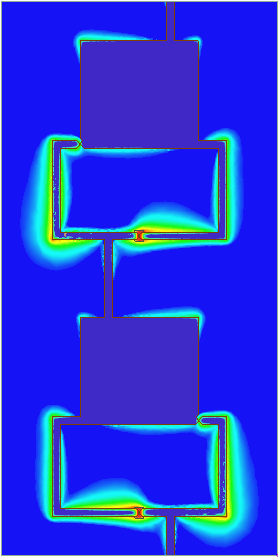} 
\label{fig:unitcell-b2}
}
~
\hspace{-0.5cm}
\subfloat[]{
\centering
\includegraphics[width=23mm]{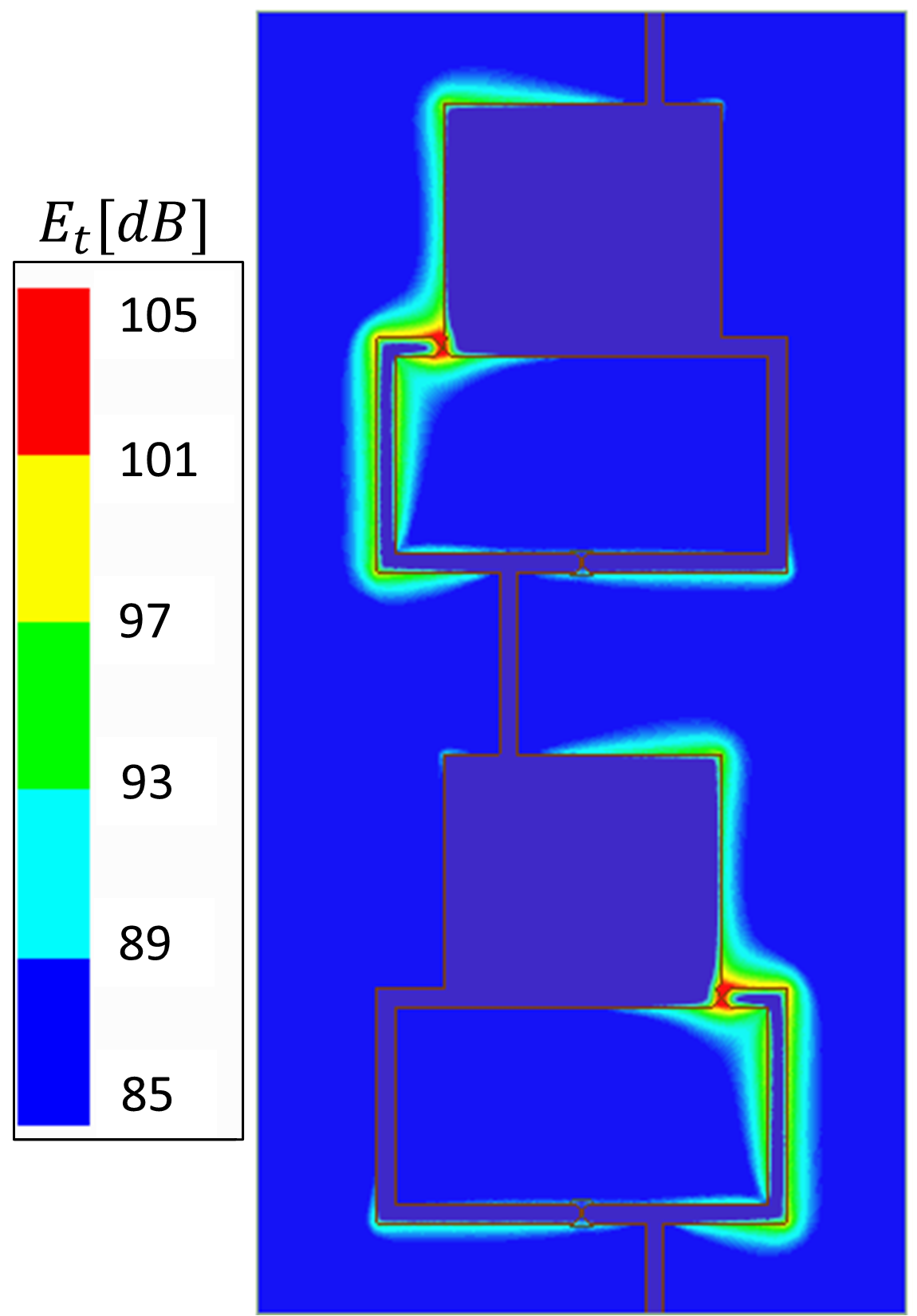} 
\label{fig:unitcell-b3}
}
~
\hspace{-0.5cm}
\subfloat[]{
\centering
\includegraphics[width=16.5mm]{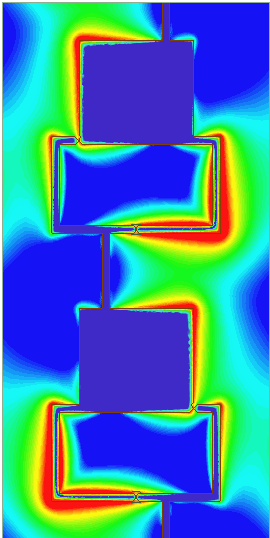} 
\label{fig:unitcell-b4}
}

\caption{Tangential electric field distribution at (a) the first resonance in the OFF state, (b) the second resonance in the OFF state, (c) the first resonance in the ON state, and (d) the second resonance in the ON state.}

\vspace{0.02cm}
\label{fig:unitcelltan}
\end{figure}

\begin{figure}[t]
    \centering

    \subfloat[]{%
        \includegraphics[width=1\linewidth]{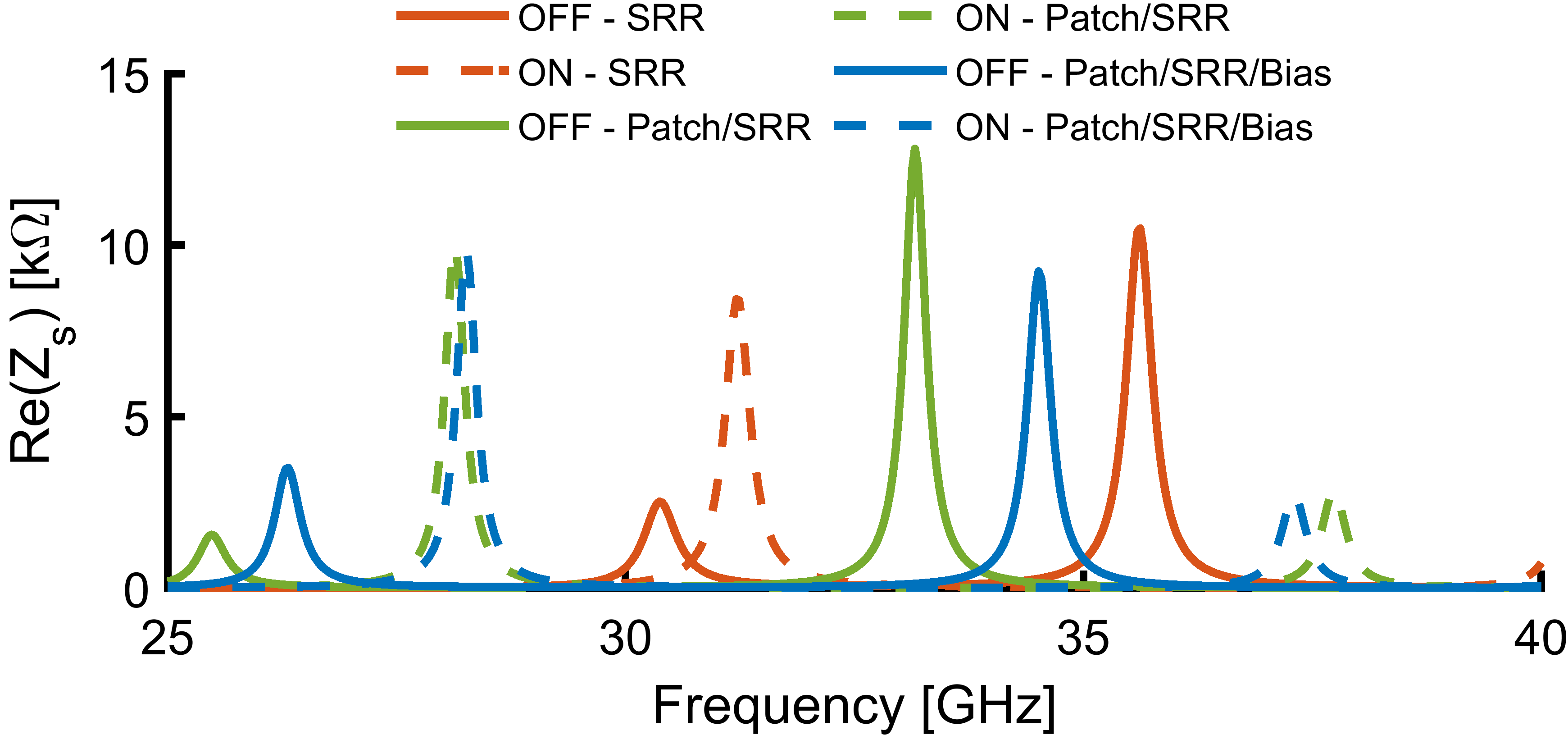}%
        \label{fig:s31}%
    }\\[-0.4ex]

    \subfloat[]{%
        \includegraphics[width=1\linewidth]{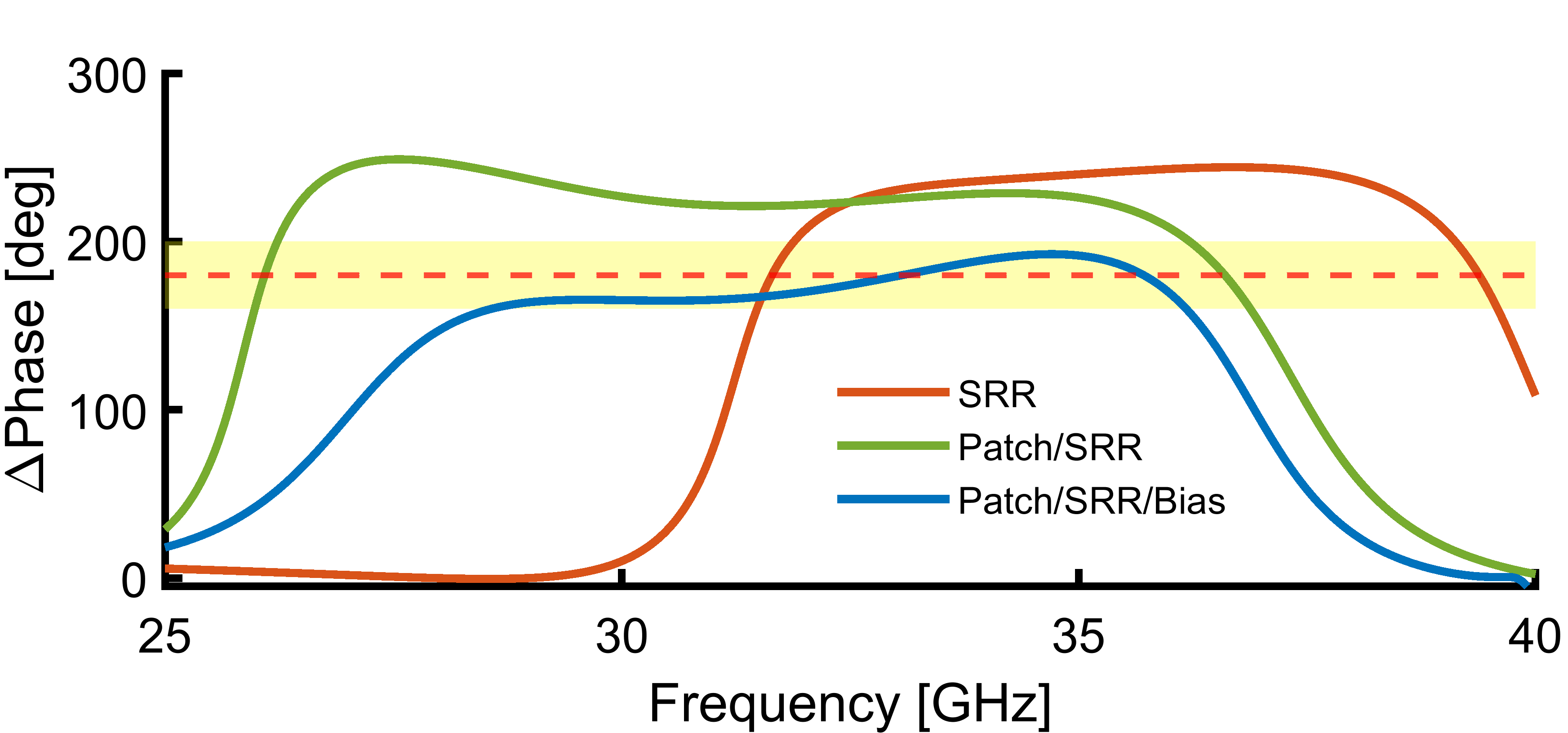}%
        \label{fig:s32}%
    }

    \caption{Impact of structural components on the (a) resonance behavior and (b) phase response of the unit cell.}
    \label{fig:s3}
\end{figure}

Fig.~\ref{fig:unitcelltan}\subref{fig:unitcell-a1}--\subref{fig:unitcell-b4}
 shows the tangential electric field distribution corresponding to the four resonances discussed, evaluated near the top surface of the unit cell. These resonances originate from the split-ring resonator (SRR) portion of the structure. When the VO$_2$ switch is turned ON, one of the SRR gaps is effectively shorted, resulting in a modified field distribution and the emergence of two resonances that are distinct from those in the OFF state.

Fig.~\ref{fig:s3} illustrates the effect of incorporating the patch element and biasing lines on the resonance behavior of the unit cell. As seen in  Fig.~\ref{fig:s3}\subref{fig:s31}, the SRR alone exhibits two distinct resonances in each switching state. Introducing the patch shifts these resonances to lower frequencies and modifies their spacing. Additionally, the inclusion of biasing lines enables further tuning of the relative frequency spacing and resonance strength, thereby allowing control over the resulting phase difference, as shown in Fig.~\ref{fig:s3}\subref{fig:s32}. This structural configuration enables optimization within a targeted frequency range and facilitates the achievement of a desired phase response.

\noindent The proposed unit cell exhibits several important features:
\begin{itemize}
    \item It achieves wideband operation with the mechanism described earlier on a single substrate with two metallic layers—of which only one is patterned—and without the need for vias, since the biasing lines are incorporated as part of the electromagnetic design.
    \item Moreover, the structure is designed such that by controlling only a negligible portion of the unit-cell area, a full $180^\circ$ phase shift is achieved.
\end{itemize}
These combined characteristics distinguish the proposed unit cell among existing RIS implementations.

In conclusion, the main objective of this work lies in the realization of a fully microfabricated RIS with seamlessly monolithically integrated VO$_2$ switches. The proposed unit cell serves as an enabling structure that allows such integration while maintaining wideband performance.

\section{Super Cell Test in a Waveguide}
\label{sec:supercell_waveguide}

In our previous work~\cite{a73}, the functionality of the designed unit cell was validated using a supercell composed of $2\times4$ sub-cells in a waveguide setup, where ideal switches (open and short circuits) were employed to emulate the switching behavior. In the present study, this analysis is extended by incorporating actual VO$_2$ switches into the same unit-cell structure to confirm that the expected phase response remains consistent. The VO$_2$ switches used are of the E-MIT type, which operate by applying a direct voltage across the terminals of the switch. When a critical electric field is reached within the junction, the device undergoes a sharp reduction in resistance—typically exceeding three orders of magnitude—which may lead to a potentially damaging inrush current.

\begin{figure}[b]
    \centering
    \includegraphics[width=0.5\textwidth]{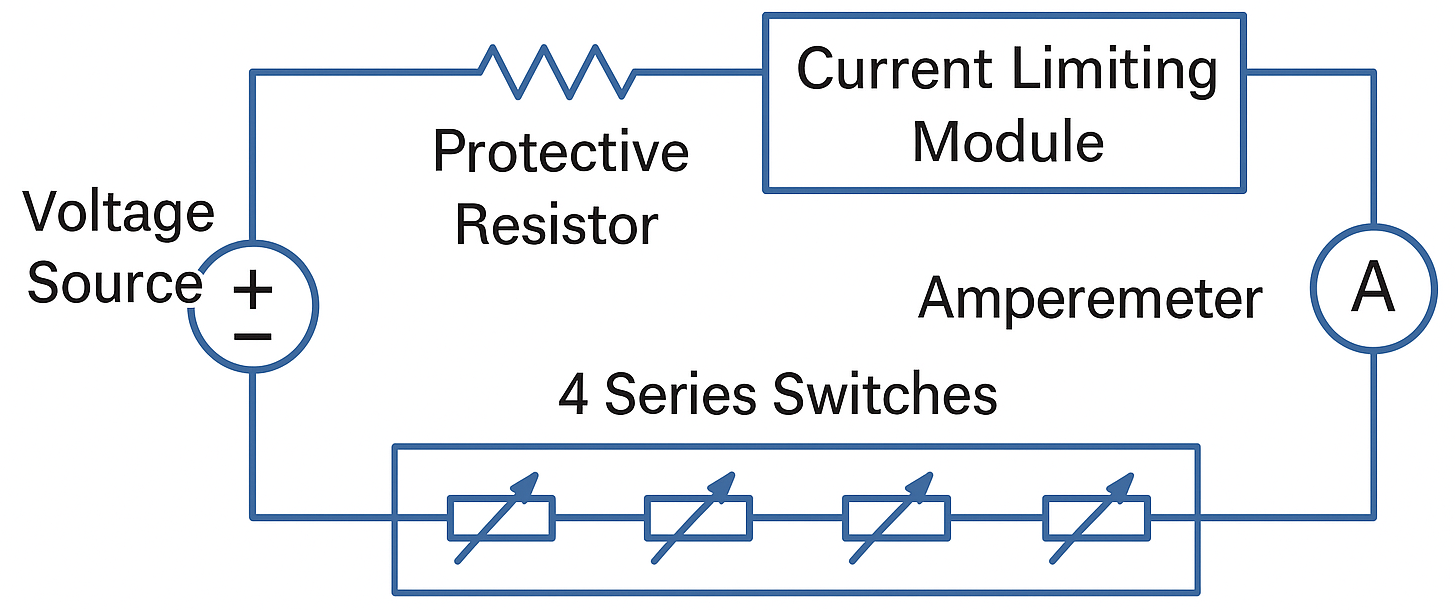}  
   \caption{Biasing circuit for each column of unit cells.}

    \label{fig:madar}
\end{figure}

To mitigate this effect and protect the device, a current limiting circuit—illustrated in Fig.~\ref{fig:madar}—was integrated into the actuation path to cap the current at 100~mA, a threshold determined to be safe for our VO$_2$ switch implementation (see Section~\ref{sec:switch_fab} for fabrication details). Additionally, a 100~$\Omega$ series resistor was included as a secondary safeguard, which can be optionally removed in practical applications. Since each column of the supercell contains four series-connected switches, a dedicated protection circuit is employed for each column to ensure independent current limiting and device safety.

\begin{figure}[t]
\centering
\vspace{-0.1cm}

\subfloat[]{
  \centering
  \includegraphics[width=60mm]{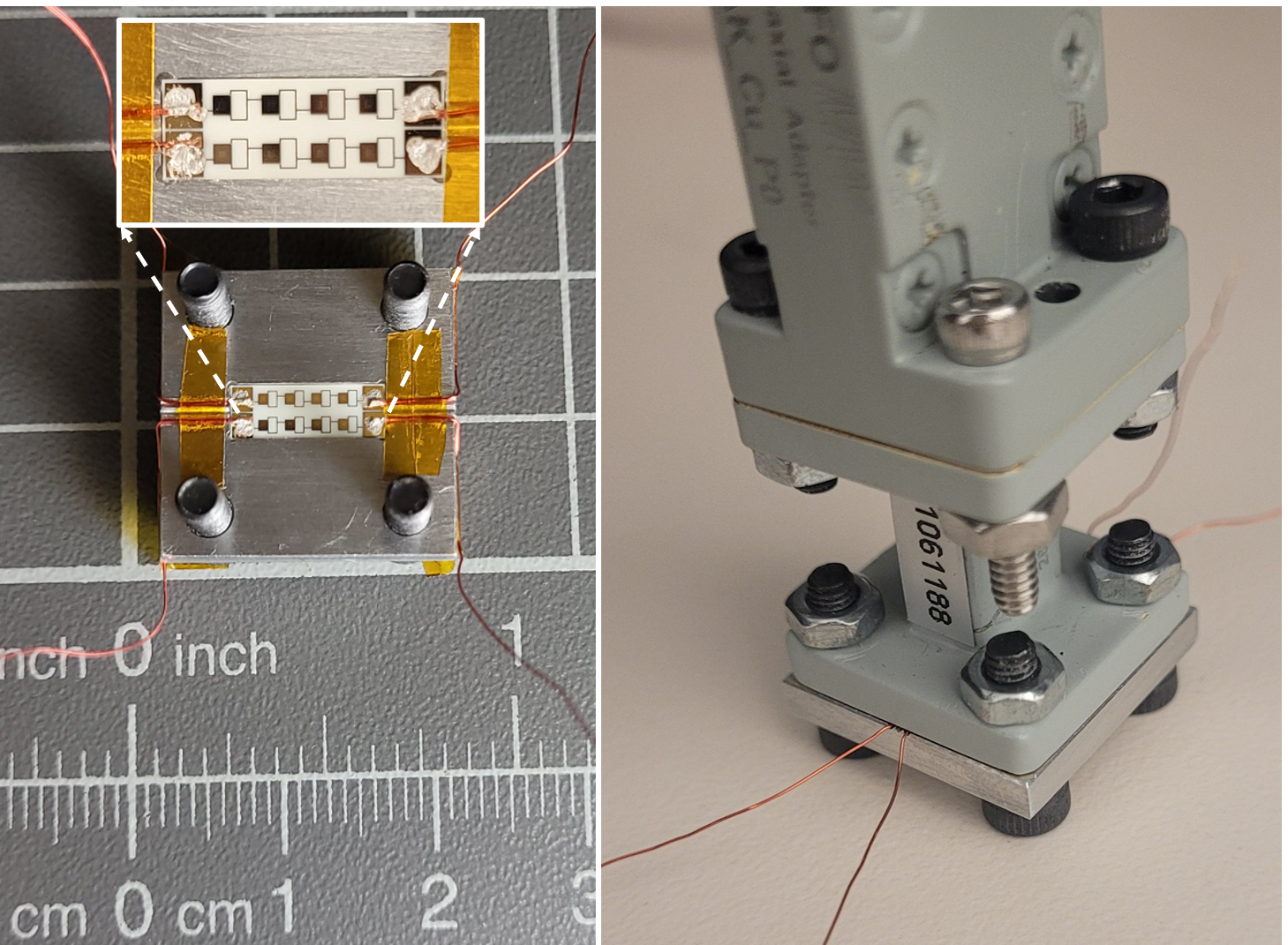} 
  \label{fig:unitcell-a11}
}

\vspace{0.02 cm} 

\subfloat[]{
  \centering
  \includegraphics[width=60mm]{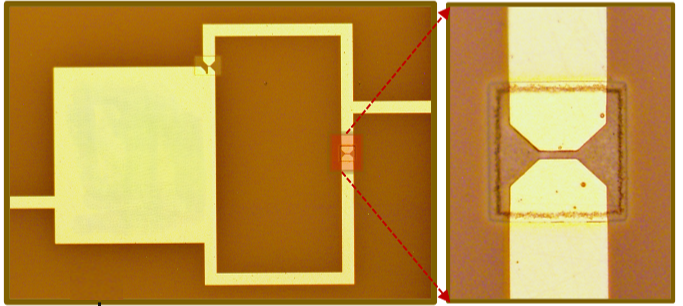} 
  \label{fig:unitcell-b11}
}

\caption{(a) Waveguide measurement setup with assembled supercell and biasing wires. 
(b) Microfabricated unit cell with integrated VO$_2$ switch.}

\label{fig:unitcell}
\vspace{0.02cm}
\end{figure}

The measurement setup closely follows the configuration used in the ideal-switch tests, with minor modifications. Specifically, grooves were introduced into the metallic holder to accommodate the biasing wires. The wires, of type 32\,AWG, are enameled for electrical isolation, sufficiently thin to preserve the structural integrity of the fixture, and can be precisely positioned onto the small DC pads using silicon-based conductive paint. 
In the final waveguide measurement setup, shown in Fig.~\ref{fig:unitcell}\subref{fig:unitcell-a11}, the waveguide aperture is aligned to expose only the supercell region, ensuring that the biasing pads and wires remain outside the aperture and do not interfere with the TE\textsubscript{10} mode of the waveguide. The biasing wires, connected to the supercell pads, are routed out of the structure to interface with the protection circuit and apply the required voltage to switch the states of the unit cells under test. Fig.~\ref{fig:unitcell}\subref{fig:unitcell-b11} shows the details of the microfabricated unit cell together with the integrated VO$_2$ switch.

\begin{figure}[b]
    \centering
    \includegraphics[width=0.5\textwidth]{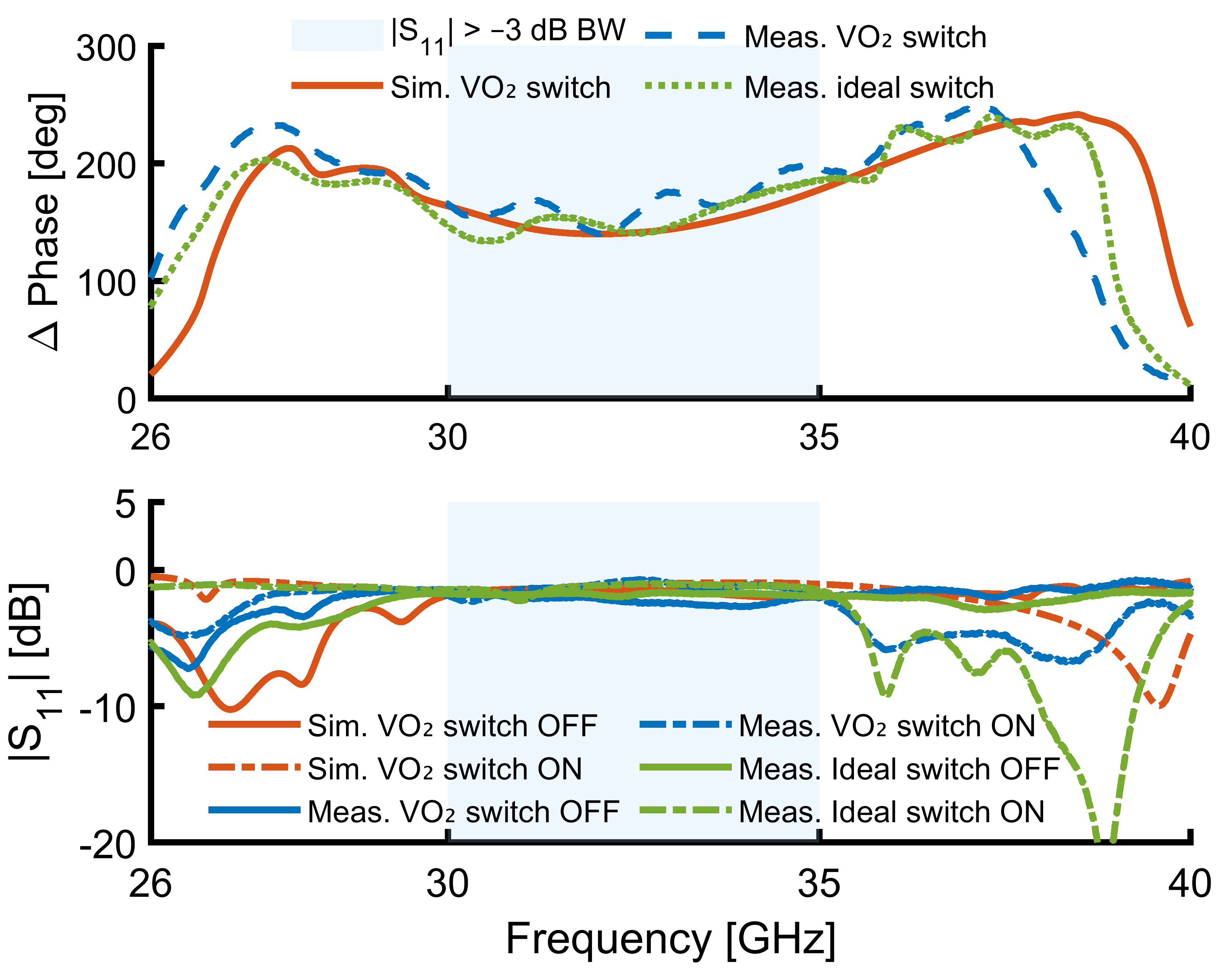}  
  \caption{Phase and amplitude response of the supercell: comparison between simulation results, waveguide measurements with ideal switches, and measurements with integrated VO$_2$ switches.}
\vspace{-0.3cm}

    \label{fig:81}
\end{figure}

Fig.~\ref{fig:81} shows the measurement results for this experiment. The phase and amplitude responses of the unit cells in the waveguide setup closely align with both full-wave simulation results and prior measurements in which ideal switches were used in place of the VO$_2$ switches. These results demonstrate that the VO$_2$ switches operate effectively as ideal switching elements within the proposed unit cell, and that their monolithic integration does not introduce observable parasitic effects that meaningfully degrade the phase or amplitude response over the measured frequency range. During the measurement, the switches were maintained in the ON state by applying a constant 20~V bias, with each column drawing approximately 75~mA of current. The resistance of each switch transitioned from approximately 5~k$\Omega$ in the OFF state to less than 3~$\Omega$ in the ON state. The highlighted zone in Fig.~\ref{fig:81} indicates the operational bandwidth where the reflection loss remains below 3~dB.

\section{RIS Design and Simulation}
\label{sec:ris_design_simulation}

The design of the proposed unit cell was thoroughly investigated, and its suitability for implementation in a \mbox{1-bit}
 reconfigurable intelligent surface (RIS) architecture was demonstrated. Building upon this foundation, our objective is to realize an RIS capable of steering reflected beams toward multiple target angles. Specifically, we consider redirection toward $\theta_r = 40^\circ$, $50^\circ$, and $60^\circ$ as representative cases. 

Under normal conditions, a conventional reflective surface redirects an incoming plane wave at an angle equal to the angle of incidence, consistent with the law of reflection. In contrast, our goal is to steer a normally incident plane wave ($\theta_i = 0^\circ$) toward a specified direction $\theta_r$ by imposing a tailored phase distribution across the RIS aperture. Based on the well-established beam-steering principle for reflective surfaces, the required surface phase profile should emulate that of a surface illuminated by a wave incident from the desired direction ($\theta_i = \theta_r$). This leads to the following expression for the necessary surface phase distribution:
\begin{equation}
    \phi_s = -k_0 d \sin{\theta_r},
\end{equation}
where $k_0$ is the free-space wavenumber and $d$ denotes the position of each unit cell along the $y$-axis of the RIS.

\begin{figure}[b]
    \centering
    \includegraphics[width=\linewidth]{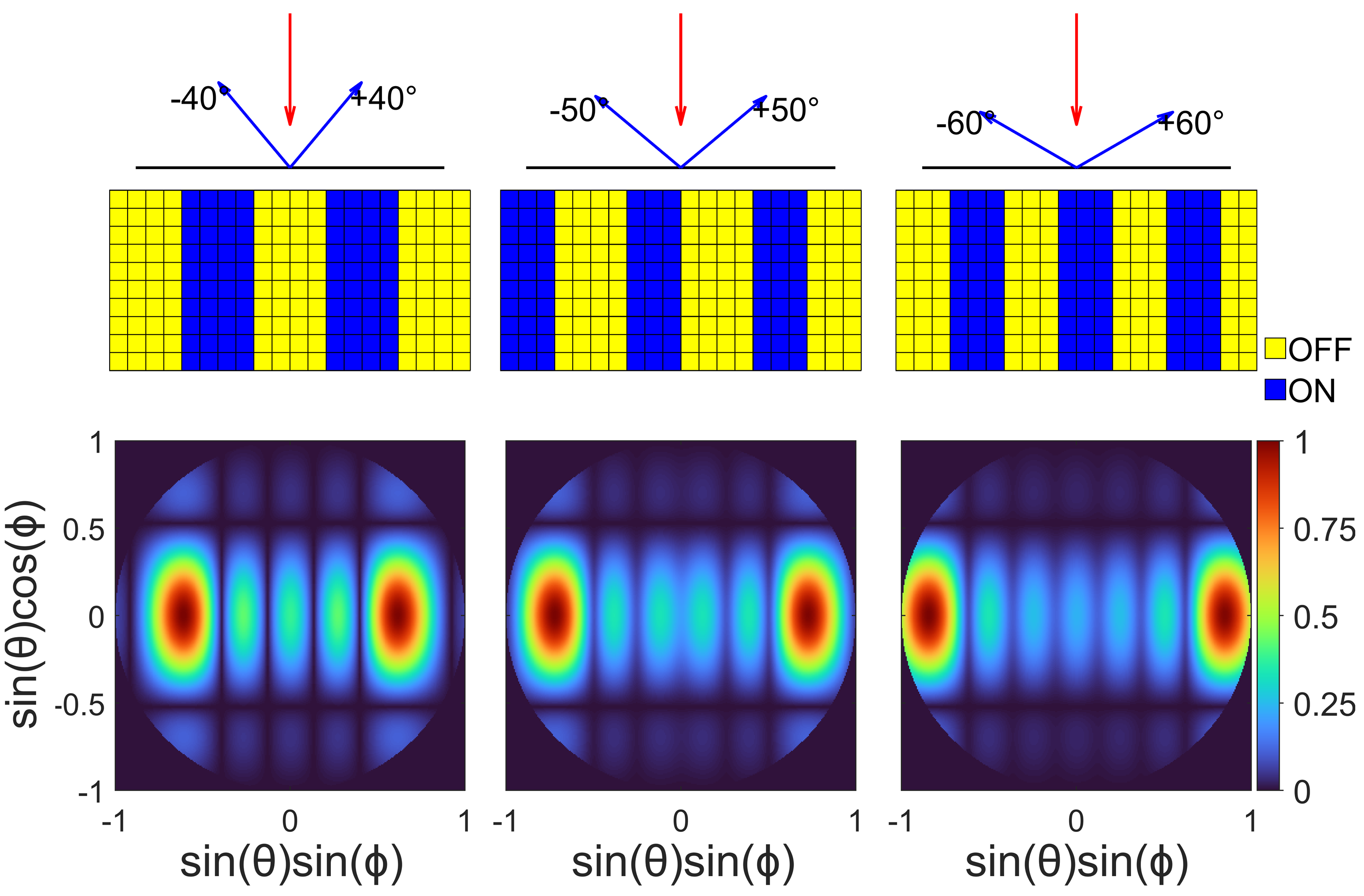} 
 \caption{Designed RIS phase patterns and corresponding normalized far-field radiation patterns for $\theta_r = \pm40^\circ$, $\pm50^\circ$, and $\pm60^\circ$, assuming ideal unit cells.}
\vspace{-0.3cm}

    \label{fig:ris_patterns_1ideal}
\end{figure}

Since the unit cell operates as a 1-bit element, the continuous phase profile $\phi_s$ must be discretized into two quantized states, as defined by the condition in (\ref{eq:quantization}). For each target angle $\theta_r$, this rule generates a specific ON/OFF pattern across the RIS aperture, forming the desired binary phase distribution. These RIS patterns are synthesized at the center frequency of the unit cell of 33~GHz, enabling efficient redirection of the incident beam toward the intended direction.

\begin{equation}
\phi_q = 
\begin{cases}
0, & 0^\circ < \phi_s < 180^\circ \\
\Delta \phi_{\text{ON--OFF}}, & 180^\circ < \phi_s < 360^\circ
\end{cases}
\label{eq:quantization}
\end{equation}

As is typical in 1-bit coded metasurfaces, the coding pattern designed to steer a beam toward an angle $\theta_{r}$ simultaneously produces another beam at $-\theta_{r}$. This symmetry in the far-field radiation pattern arises because the 1-bit coded RIS cannot distinguish between the $+\theta_{r}$ and $-\theta_{r}$ directions when illuminated by a plane wave. The additional beam appearing at the mirrored direction corresponds to the so-called quantization lobe.


In each unit cell, the reflected $E_y$ field can be expressed as $A_{\text{cell}} e^{j \phi_{\text{cell}}}$, where the reflection amplitude $A_{\text{cell}}$ and phase $\phi_{\text{cell}}$ depend on the switching state (ON or OFF). Specifically, $A_{\text{cell}} \in \{A_{\text{on}}, A_{\text{off}}\}$ and $\phi_{\text{cell}} \in \{\phi_{\text{on}}, \phi_{\text{off}}\}$.

\begin{figure}[t]
    \centering

    \includegraphics[width=\linewidth]{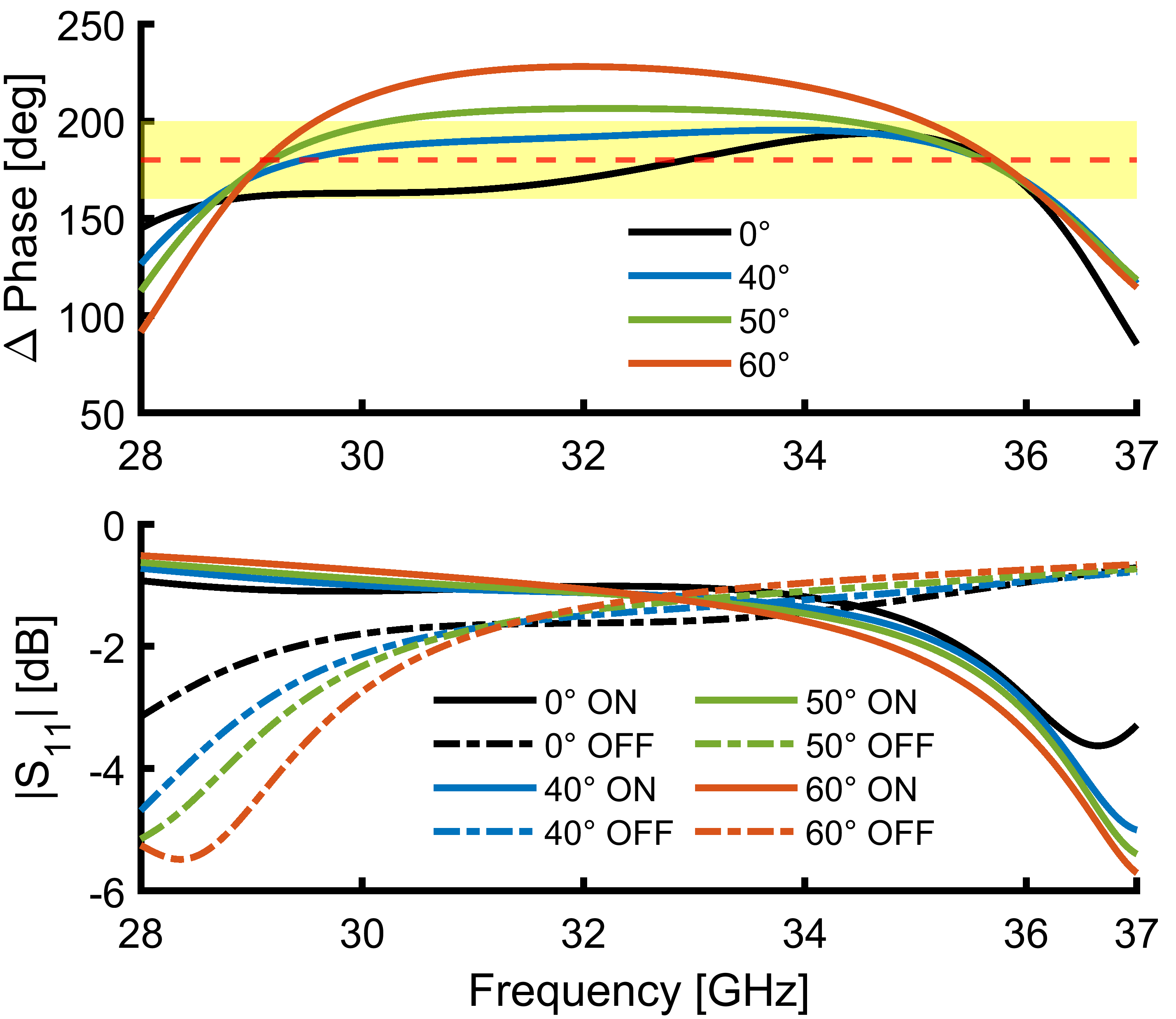}
   \caption{Simulated frequency-dependent reflection amplitude and phase difference of the unit cell under periodic boundary conditions for various incidence/reflection angles.}
\vspace{-0.3cm}
    \label{fig:unitcell_sparams_a}
\end{figure}

In the ideal case, the reflection amplitudes for both states are equal and unity, i.e., $A_{\text{on}} = A_{\text{off}} = 1$, and the phase difference between the ON and OFF states is exactly $180^\circ$, such that $\phi_{\text{on}} - \phi_{\text{off}} = \pi$.

\begin{figure}[b]
    \centering
    \includegraphics[width=0.8\linewidth]{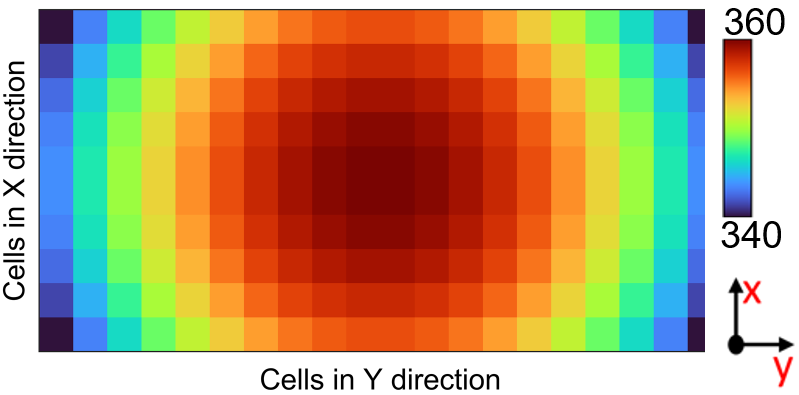}  
   \caption{Incident phase distribution $\phi_i$ from a point source located 29\,cm from the RIS aperture.}

    \label{fig:unitcell_sparams_b}
\end{figure}

\begin{figure*}[t]
\centering
\vspace{-0.3cm}

\subfloat[]{
\includegraphics[width=0.315\textwidth]{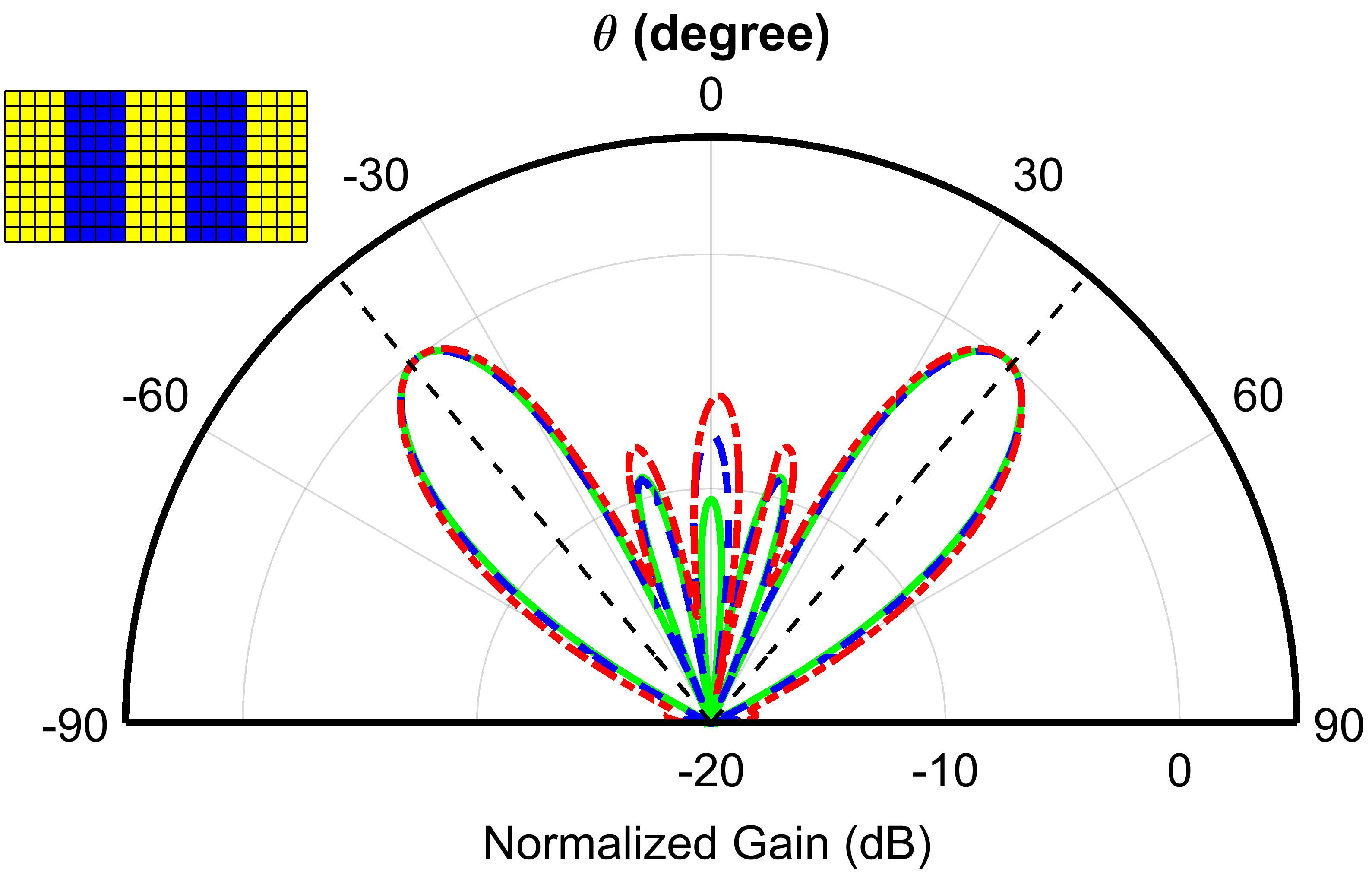}
\label{fig:fig1}
}
\hfill
\subfloat[]{
\includegraphics[width=0.315\textwidth]{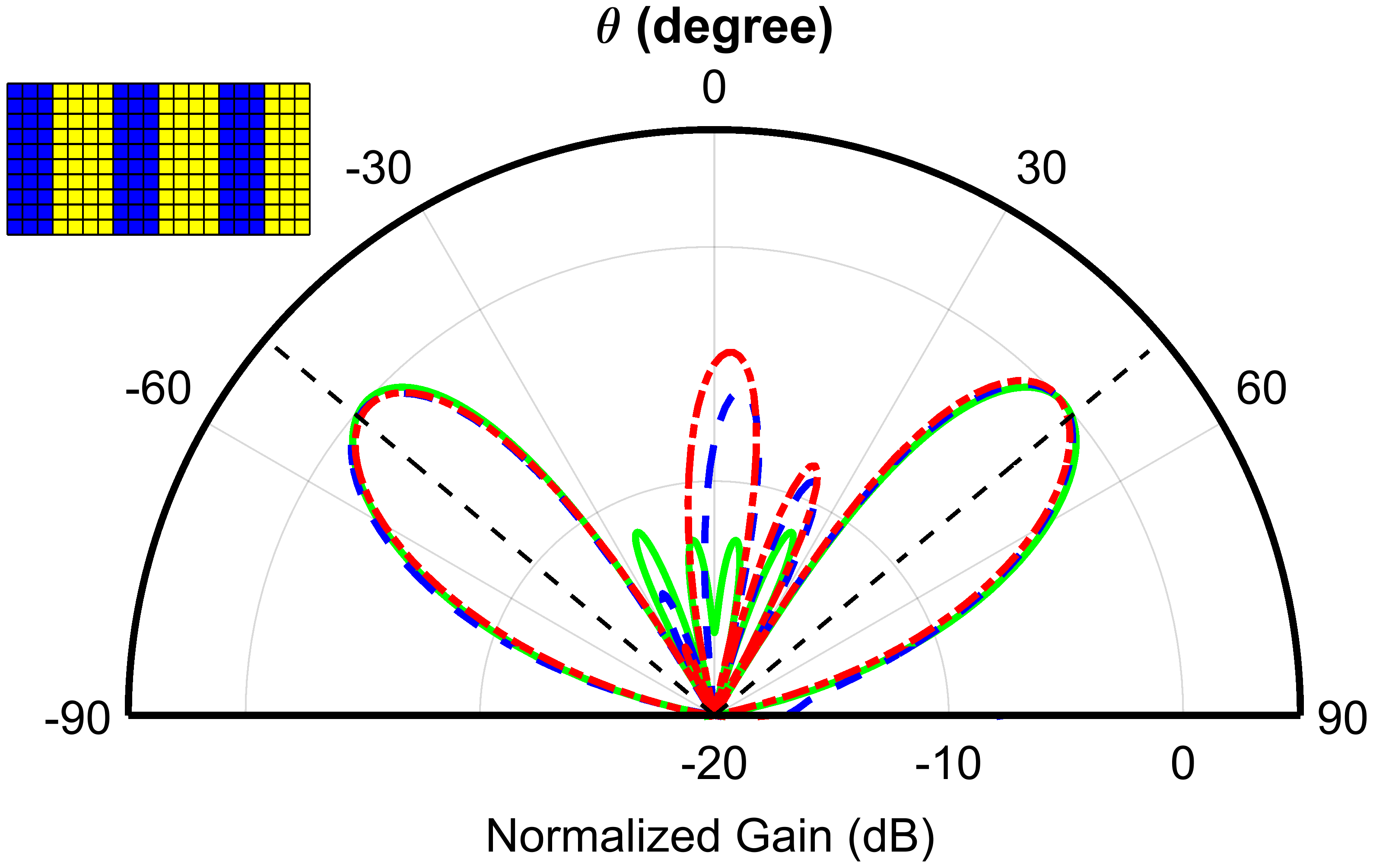}
\label{fig:fig2}
}
\hfill
\subfloat[]{
\includegraphics[width=0.315\textwidth]{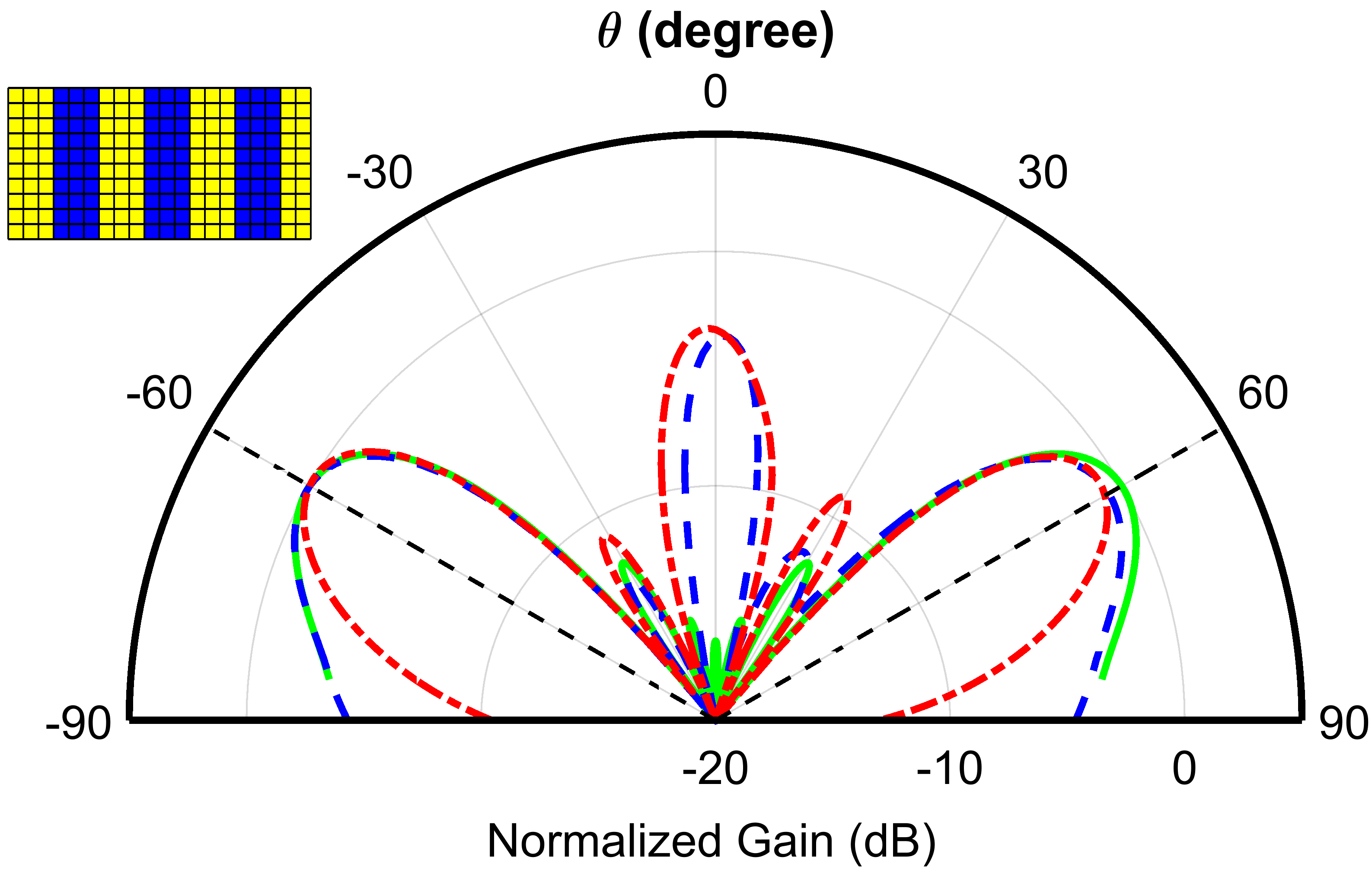}
\label{fig:fig3}
}

\vspace{0.2cm}

\caption{Normalized gain results for RIS patterns designed to reflect beams at (a)~$\pm40^\circ$, (b)~$\pm50^\circ$, and (c)~$\pm60^\circ$ at 33\,GHz.
Legend: 
\protect\tikz[baseline=-0.6ex]\protect\draw[green, thick] (0,0)--(0.6,0); Numerical (ideal unit cell), 
\protect\tikz[baseline=-0.6ex]\protect\draw[blue, thick, dash pattern=on 4pt off 2pt] (0,0)--(0.6,0); Semi full-wave, 
\protect\tikz[baseline=-0.6ex]\protect\draw[red, thick, dash pattern=on 1pt off 1pt] (0,0)--(0.6,0); Full-wave simulation.}
\label{fig:single-row}
\vspace{-0.3cm}
\end{figure*}
However, in practical implementations, the reflection amplitudes deviate from unity and may differ between the two switching states. Likewise, the phase difference is not exactly $\pi$ and can vary across the operational bandwidth, depending on the design characteristics of the 
unit cell. With the reflection amplitude and phase defined across the entire RIS surface, the near-field distribution of the reflected electric field—specifically the $E_y$ component, given the RIS operates under $y$-polarized excitation—can be constructed. The far-field radiation pattern is then obtained by applying a spatial Fourier transform to this tangential field distribution over the RIS aperture, as expressed by:

\begin{align}
\vec{E}_{\text{ff}} =\; & \frac{e^{-jkr}}{2\pi r} \cdot jk \left( \hat{\theta} \sin\phi + \hat{\phi} \cos\phi \cos\theta \right) \nonumber \\
& \times \int_{-\frac{NP}{2}}^{\frac{NP}{2}} \int_{-\frac{MP}{2}}^{\frac{MP}{2}} 
E_y(x, y) e^{jk_0(ux + vy)} \, dx \, dy,
\end{align}

\noindent
where $u = \sin\theta \cos\phi$, $v = \sin\theta \sin\phi$, and $P$ is the unit-cell size. The parameters $M$ and $N$ represent the number of unit cells along the $x$- and $y$-directions, respectively.

Since the $E_y$ field is inherently discretized over the RIS surface—taking quantized amplitude and phase values determined by the binary states of the unit cells—the spatial Fourier transform must be computed numerically. This is efficiently achieved using the Fast Fourier Transform (FFT), applied to the discrete field distribution defined by the RIS configuration. The numerical implementation of this Fourier-based far-field extraction follows the approach described in~\cite{a74}.

The designed RIS patterns and their corresponding far-field radiation responses for three different beam steering angles—assuming ideal unit cells—are presented in Fig.~\ref{fig:ris_patterns_1ideal}. In this ideal case, each unit cell is modeled with constant and quantized reflection phase and unit-amplitude response over the entire frequency band, as previously described.

To enhance the realism of the simulation setup, the radiation patterns are recalculated using frequency-dependent reflection coefficients extracted from full-wave simulations of the unit cells. This refined approach accounts for variations in the reflection angle and accurately captures the frequency-dependent amplitude and phase response of the unit cells when illuminated at different incidence angles under periodic boundary conditions. The corresponding simulation results for various angles of incidence and reflection are presented in Fig.~\ref{fig:unitcell_sparams_a}.

Furthermore, instead of assuming ideal plane-wave excitation, the illumination source is modeled as a point source located 29~cm from the RIS aperture (approximately 30 wavelengths away), introducing a mild phase gradient across the surface. This incident phase profile, denoted as $\phi_i$, is superimposed on the designed reflection phase $\phi_q$ to provide a more physically realistic excitation in the numerical model. The resulting excitation phase distribution is shown in Fig.~\ref{fig:unitcell_sparams_b}, where the maximum phase difference between the central and corner unit cells is approximately $22^\circ$. This demonstrates that a source placed 29~cm away from the RIS offers a very good approximation of plane-wave illumination.

The resulting radiation patterns are then compared against those obtained from both the idealized configuration and a full-wave electromagnetic simulation of the complete $10 \times 20$ RIS structure in HFSS, illuminated by a point source positioned 29~cm from the aperture. The normalized $E_\theta$ far-field patterns in the $yz$-plane (steering plane) for each configuration are presented in Fig.~\ref{fig:single-row}.

\begin{figure}[b]
    \centering
    \includegraphics[width=0.5\textwidth]{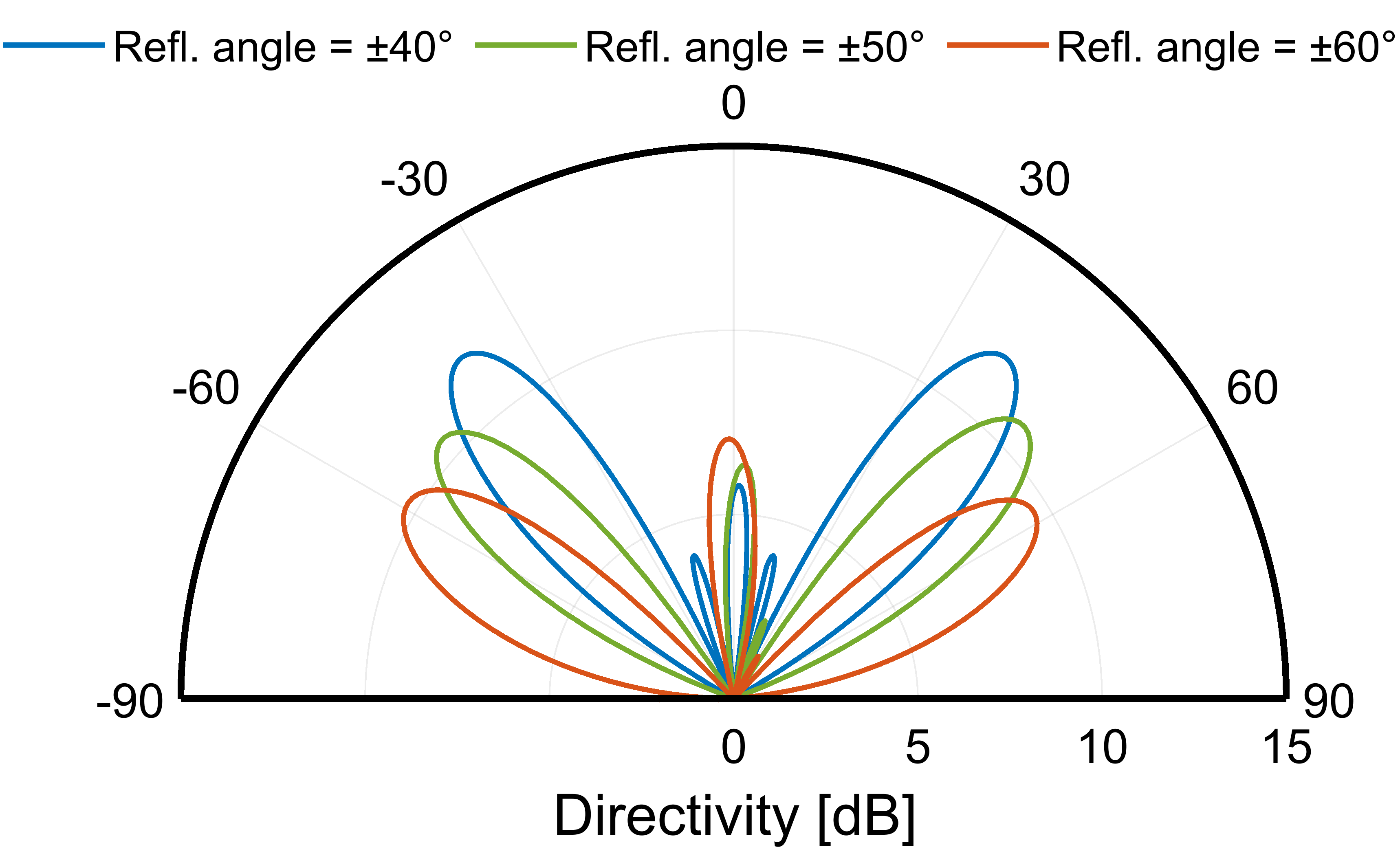}  
 \caption{Simulated far-field directivity patterns of the RIS in the $yz$-plane.}

    \label{fig:qq1}
\end{figure}

As shown in Fig.~\ref{fig:single-row}\subref{fig:fig1}--\subref{fig:fig3}, the designed patterns successfully redirect the beam toward $40^\circ$, $50^\circ$, and $60^\circ$, respectively. Each subfigure presents three scenarios: numerical calculation with an ideal unit cell, semi-full-wave results using a non-ideal unit cell, and full-wave simulation using HFSS. These scenarios, previously described in detail, consistently identify the main beam direction for all three angles, indicating strong agreement between the methods. Notably, the semi-full-wave approach closely follows the full-wave simulation, not only in the main beam direction but also in capturing the sidelobe characteristics. The sidelobes become more pronounced at $50^\circ$ and $60^\circ$, where the unit cells operate near or beyond their optimal phase change range of $160^\circ$ to $200^\circ$, resulting in elevated sidelobe levels.

This study confirms the validity of using single-unit-cell full-wave data within numerical formulations to generate far-field radiation patterns comparable to those obtained through complete full-wave simulations, offering both computational efficiency and design accuracy. This approach is later employed for channel modeling and estimating the impact of the RIS on received power ratios.

Fig.~\ref{fig:qq1} illustrates the directivity of each RIS pattern in the steering ($yz$) plane, derived from full-wave simulations. As the steering angle increases from $40^\circ$ to $60^\circ$, the main lobe directivity decreases by approximately 2\,dB. Nonetheless, the directivity remains above 10\,dB across all cases, indicating that efficient beam steering is maintained even at higher deflection angles.

The observed directivity levels for the designed RIS are consistent with expectations, given the physical size, phase resolution, and steering constraints of the structure. The implemented RIS has dimensions of 3.4\,cm~$\times$~1.7\,cm, corresponding to approximately $3.8\lambda \times 1.9\lambda$ at the mid-band operating frequency of 33\,GHz. This limited aperture inherently restricts the maximum achievable directivity, which is further impacted by the use of 1-bit phase quantization and the non-ideal phase response of the unit cells. Additionally, steering to wide angles such as $\pm50^\circ$ and $\pm60^\circ$ introduces aperture projection losses due to the cosine effect, reducing the effective aperture and contributing to lower directivity.

Despite these limitations, the RIS demonstrates focused beams with directivity levels of 10--12\,dB, aligning well with theoretical expectations for compact, practical RIS structures. Importantly, this $10\times20$ element RIS can serve as a modular building block for larger implementations. By scaling the aperture while maintaining the same unit-cell design, the overall directivity can be significantly improved—since a fourfold increase in physical area theoretically yields a 6\,dB gain in directivity—enabling more efficient beam steering and broader coverage in future applications.

\noindent\textbf{Quantization and scattering considerations:} 
For clarity, quantization lobes are explicitly distinguished from specular reflection. As discussed earlier, the unwanted mirror-like lobe observed in Fig.~\ref{fig:single-row} originates from the 1-bit phase quantization symmetry (i.e., a quantization lobe). One reported mitigation approach~\cite{a79} suppresses quantization lobes by slightly misaligning adjacent RIS tiles in the perpendicular direction—an approach that is compatible with the present modular $10\times20$ tile and can be readily extended to form larger apertures.

\begin{table}[b]
\centering
\renewcommand{\arraystretch}{1.3}
\setlength{\tabcolsep}{8pt}

\caption{Optimized geometric parameters of the unit cell}
\label{tab:selected_params}

\begin{tabular}{|>{\centering\arraybackslash}m{1.3cm}
                |>{\centering\arraybackslash}m{1.1cm}
                |>{\centering\arraybackslash}m{1.1cm}
                |>{\centering\arraybackslash}m{1.1cm}
                |>{\centering\arraybackslash}m{1.1cm}|}
\hline
\textbf{Parameter} & w1 & w2 & w4 & w5 \\
\hline
\textbf{Value (µm)} & 1128.5 & 422 & 447.5 & 445.6 \\
\hline
\end{tabular}
\end{table}

\begin{figure}[t]
    \centering

    \subfloat[]{%
        \includegraphics[width=1\linewidth]{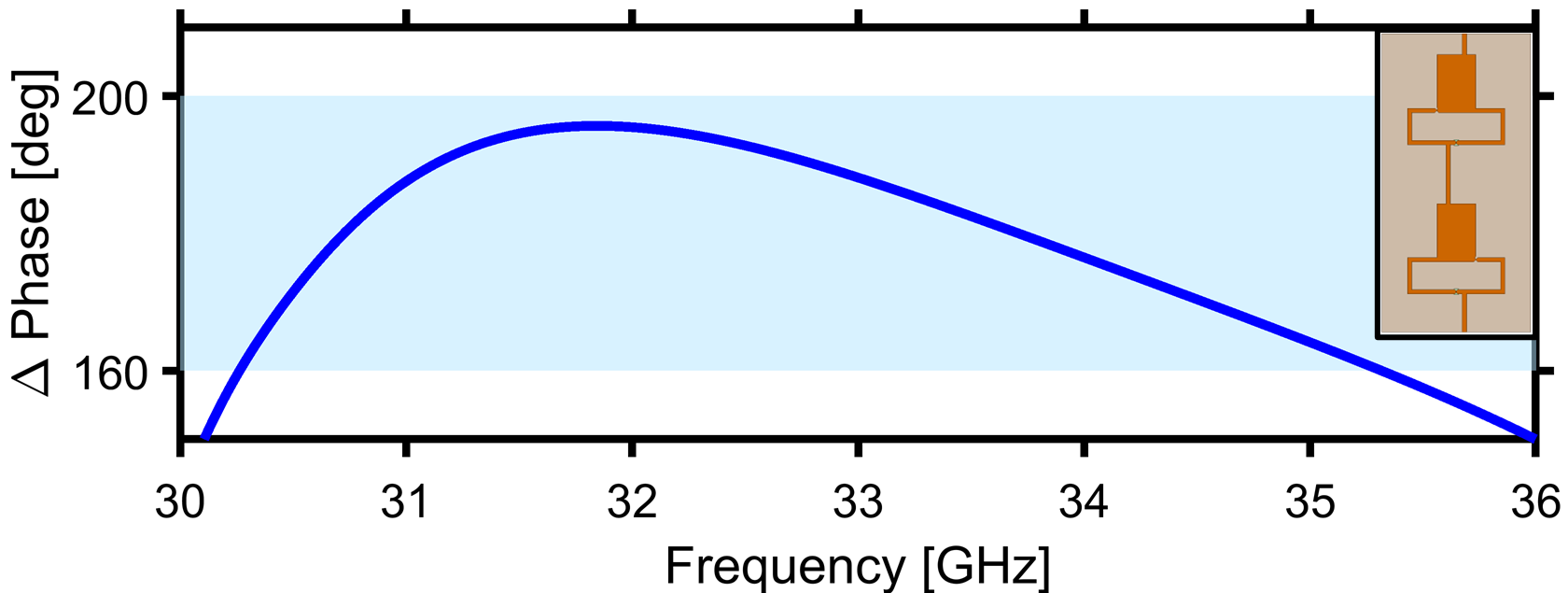}%
        \label{fig:s3q1-a}%
    }\\[-0.1ex]

    \subfloat[]{%
        \includegraphics[width=1\linewidth]{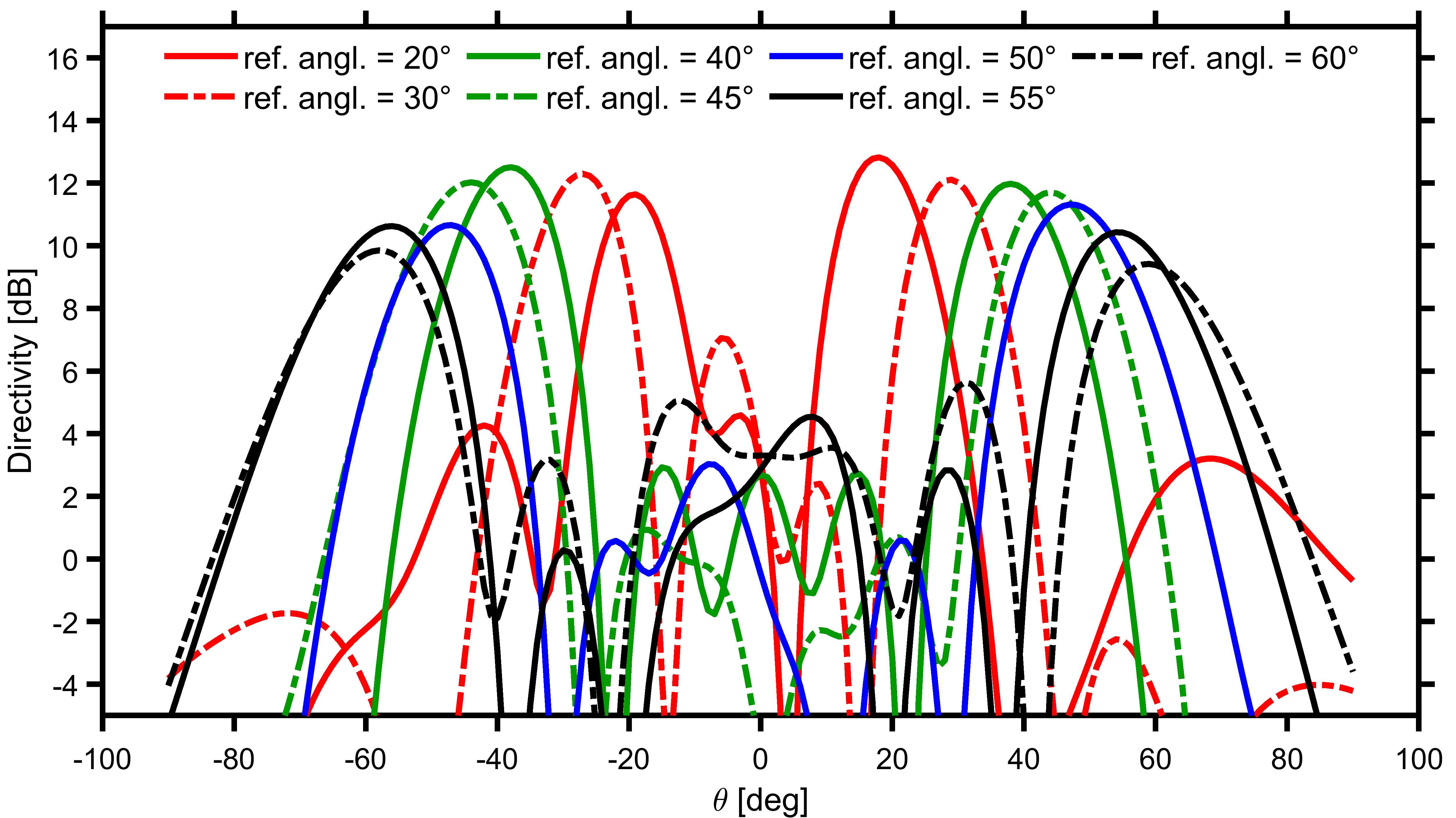}%
        \label{fig:s3q1-b}%
    }

    \caption{(a) Phase difference between the ON and OFF states when the unit cell is optimized for a reflection angle of $50^\circ$ (the inset shows the optimized unit-cell configuration).
    (b) HFSS-simulated directivity patterns for different reflection angles after unit-cell optimization.}
    \label{fig:s3q1}
\end{figure}

By contrast, the central lobe corresponds to specular reflection, which persists because the unit-cell phase contrast deviates more from the ideal $180^\circ$ as the reflection angle increases. By adjusting the unit-cell parameters, as listed in Table~\ref{tab:selected_params}, while keeping the other parameters unchanged, the relative surface resonances reposition and thereby align the ON/OFF phase contrast within $160^\circ$–$200^\circ$ for a high reflection angle of $50^\circ$. As shown in Fig.~\ref{fig:s3q1}\subref{fig:s3q1-a}, at the expense of a slightly reduced bandwidth (31–35\,GHz), a better phase contrast is achieved at higher reflection angles. For this new set of parameters, the resulting directivity patterns at the mid-band frequency for reflection angles between $20^\circ$–$60^\circ$, shown in Fig.~\ref{fig:s3q1}\subref{fig:s3q1-b}, confirm that this optimization suppresses the specular component by approximately 8\,dB below the main lobe at $50^\circ$, compared to 4\,dB before the adjustment. This optimization also reduces the specular reflection at other high reflection angles. These results demonstrate that even without array-level optimization, unit-cell optimization alone can effectively mitigate specular reflection.  

Although scattering control (quantization-lobe suppression and specular-reflection reduction) is not the primary focus of this work, these results and discussions clarify that established methods are readily available and can be employed when needed. For the remainder of this work, the primary unit-cell design parameters are used to achieve the best bandwidth for directional gain enhancement of the RIS.

\section{RIS Fabrication and Measurement Setup}
\label{sec:ris_fabrication_measurement}

\subsection*{A) Microfabrication Process and Biasing Method}
\begin{figure}[t]
    \centering

    \subfloat[]{%
        \includegraphics[width=0.85\linewidth]{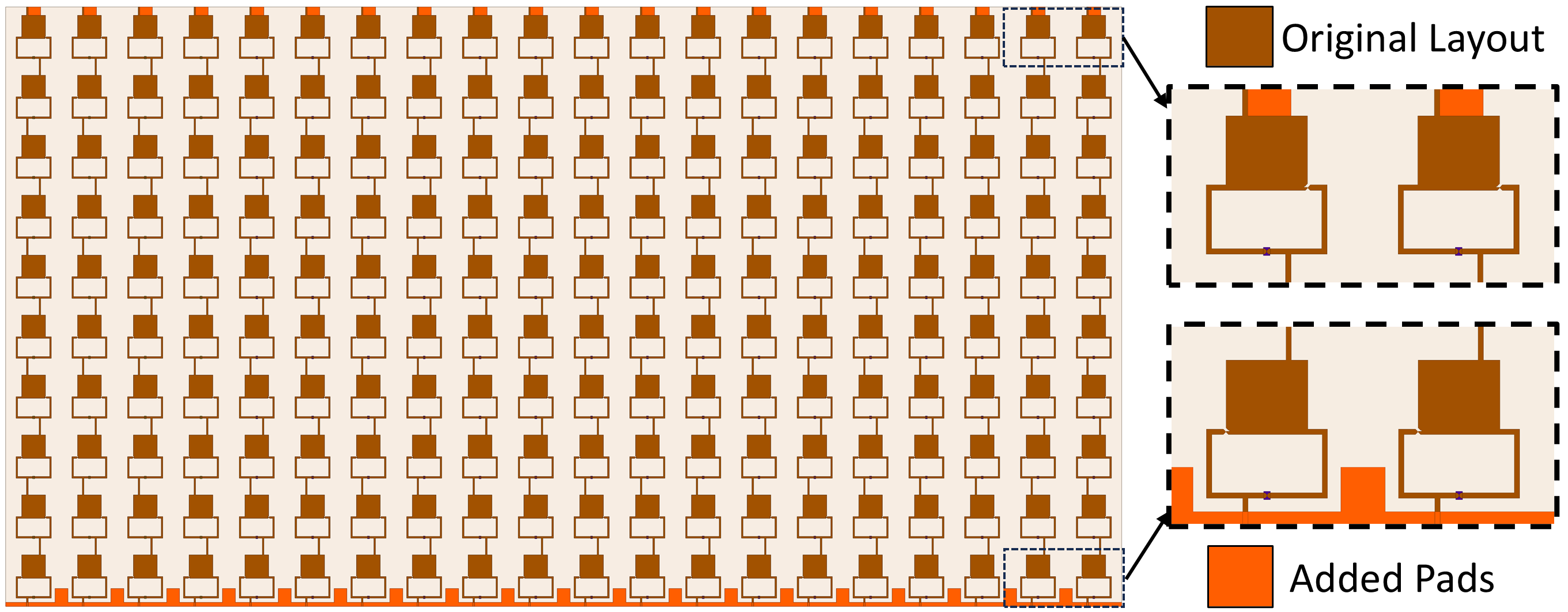}%
        \label{fig:ris_layout-a}%
    }\\[-0.1ex]

    \subfloat[]{%
        \includegraphics[width=0.85\linewidth]{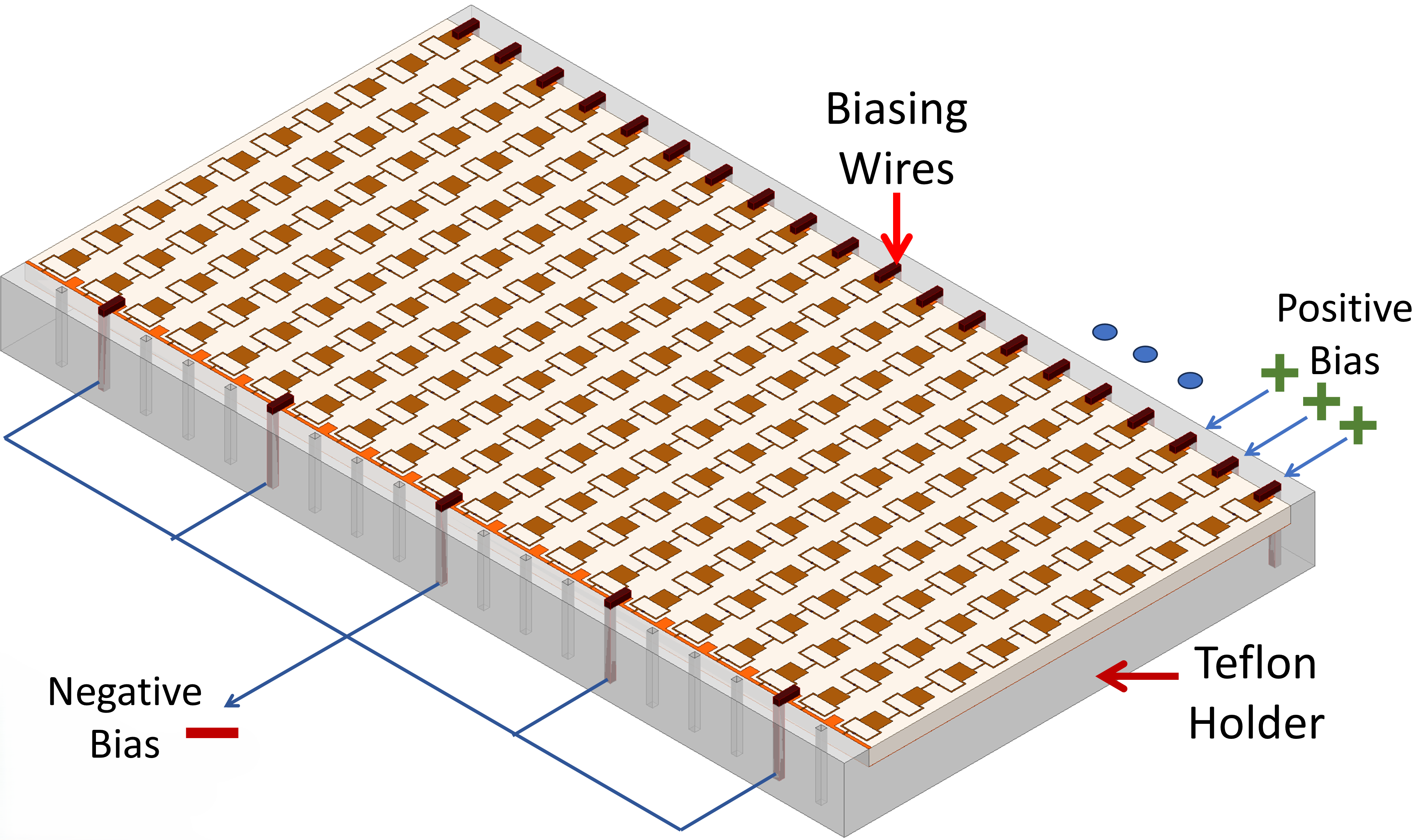}%
        \label{fig:ris_layout-b}%
    }\\[-0.01ex]

    \subfloat[]{%
        \includegraphics[width=0.85\linewidth]{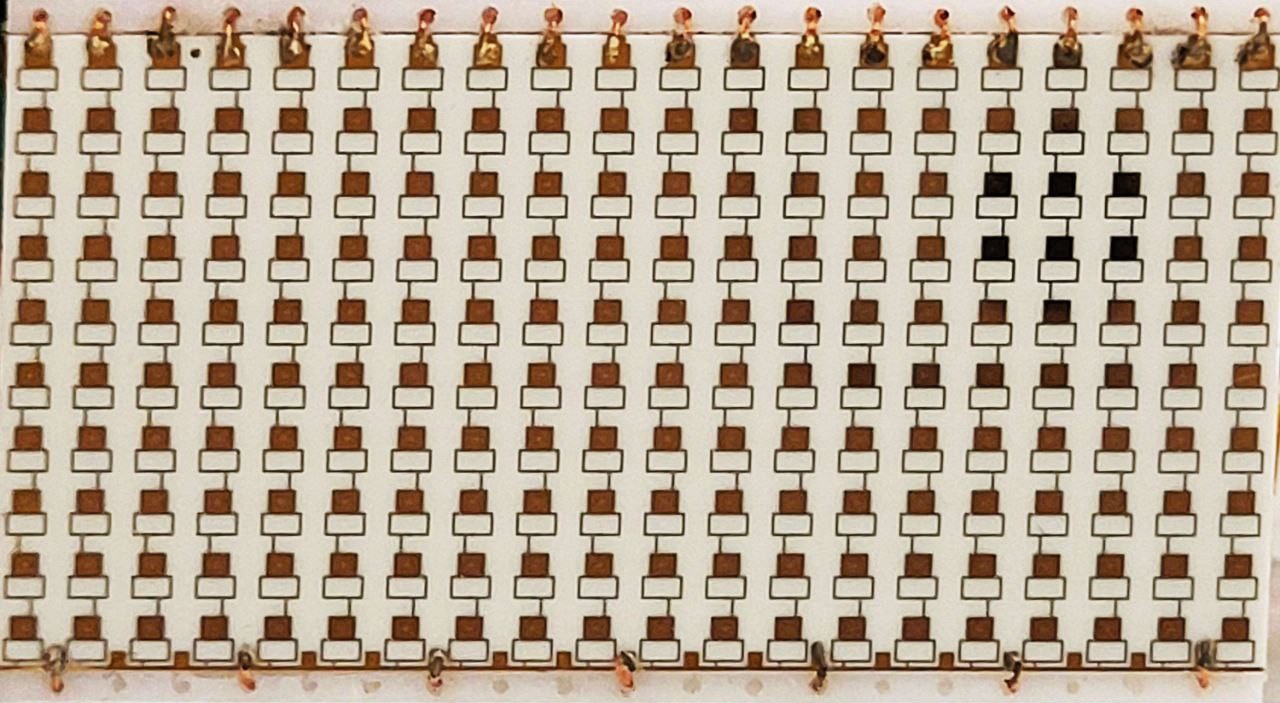}%
        \label{fig:ris_layout-c}%
    }

    \caption{(a) RIS layout modification; (b) Teflon holder and biasing configuration; (c) Microfabricated array with connected biasing wires.}
    \label{fig:ris_layout}
    \vspace{-0.3cm}
\end{figure}

The general design of the RIS, consisting of $20$ columns and $10$ rows of the proposed sub-cells, is shown in Fig.~\ref{fig:ris_layout}\subref{fig:ris_layout-a}. To enable column-wise reconfigurability, this layout is modified by incorporating additional features into the cells located along the top and bottom edges of the surface to facilitate the integration of biasing wires. These added pads are designed to be small enough to minimize their impact on the RIS response while remaining large enough to allow for reliable electrical connections. Since each column contains ten series-connected E-MIT VO$_2$ switches that must be simultaneously activated, two electrical pads per column are ideally required—one at the top (positive bias) and one at the bottom (negative bias). To simplify the assembly, the bottom pads are electrically connected across the bottom edge of the RIS to form a common node, thereby reducing the number of wires required on the bottom side. In contrast, the top pads remain electrically isolated from one another to allow individual connections to a dedicated current-limiting protection circuit, thus enabling safe and independent biasing of each column.

To implement the biasing wire connections, a custom-designed Teflon-based stand was fabricated using CNC machining, featuring a recessed region precisely matching the footprint of the RIS surface to allow it to sit securely in place. As shown schematically in Fig.~\ref{fig:ris_layout}\subref{fig:ris_layout-b}, the stand includes 20 holes along both the top and bottom edges. These openings are used to route the wires for establishing the positive and negative DC biasing terminals.

To establish electrical connections with the micron-sized DC pads, 32~AWG enameled copper wire was used—the same type previously employed in the unit-cell waveguide testing setup. This wire was selected for its thin profile and high flexibility. The enamel coating provides insulation, enabling close routing of multiple wires without risk of shorting, while maintaining a compact form factor. For accurate mechanical alignment, the wires were threaded through micro-scale holes in the Teflon structure, ensuring consistent positioning relative to the DC pads. The flexibility of the wire allows it to bend onto the pads with minimal mechanical stress. Electrical contact was achieved by carefully applying silver epoxy at the contact points under a microscope, followed by thermal curing in an oven to ensure precise and stable connections. The microfabricated RIS, along with its Teflon holder and electrical connections, is shown in Fig.~\ref{fig:ris_layout}\subref{fig:ris_layout-c}.

\begin{figure}[t]
    \centering
    \includegraphics[width=0.9\linewidth]{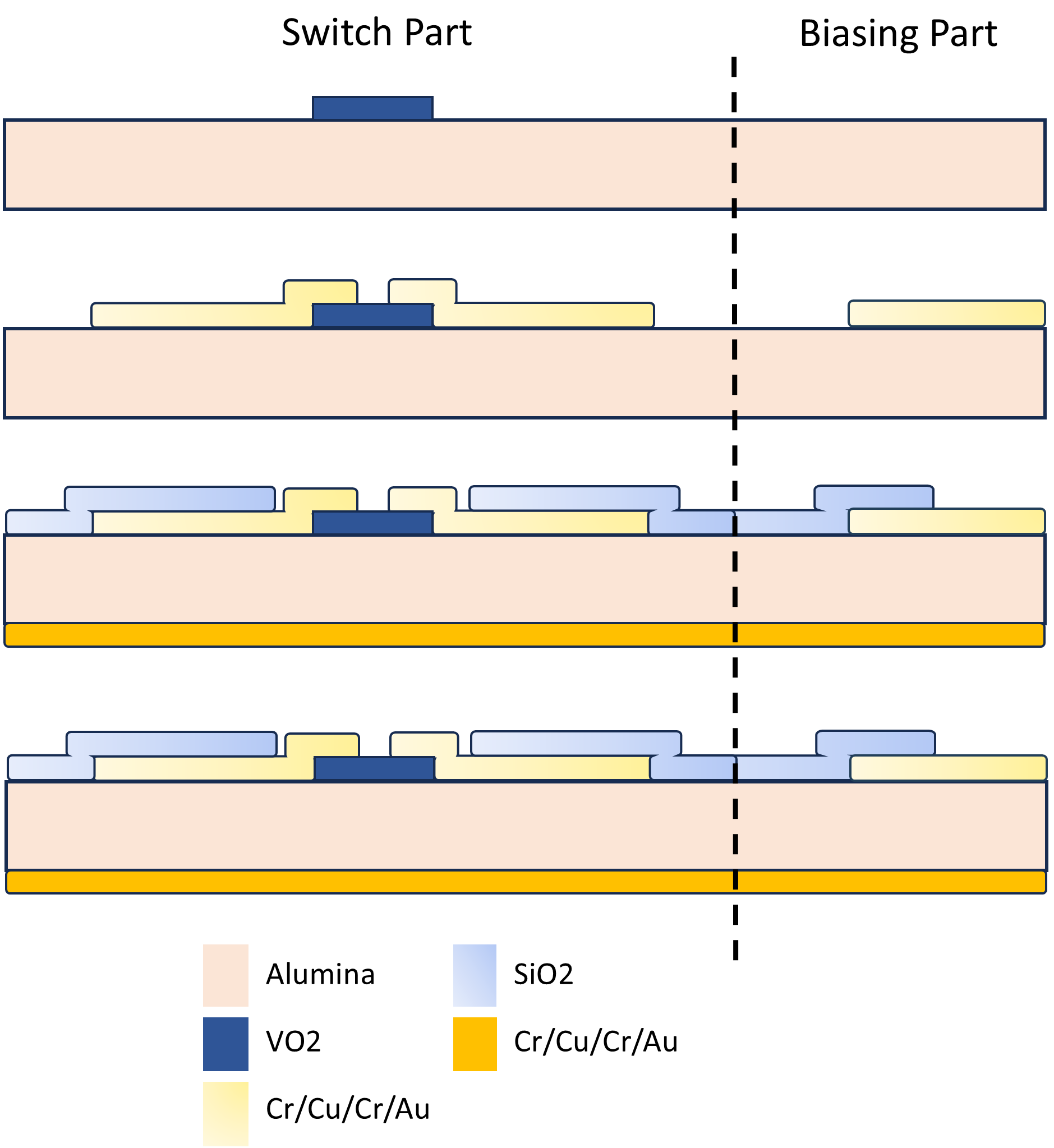} 
    \caption{Microfabrication Process}
    \label{fig:ris_patterns_ideal}
    \vspace{-0.3cm}
\end{figure}

The complete microfabrication procedure used to realize this design is summarized in Fig.~\ref{fig:ris_patterns_ideal}. The process begins with the deposition of a 350\,nm-thick VO\textsubscript{2} (vanadium dioxide) layer on an alumina substrate using the pulsed laser deposition (PLD) technique. This layer is subsequently patterned through a wet etching process utilizing a chrome-specific etchant. Next, a multilayer metal stack comprising Cr/Cu/Cr/Au is deposited via electron beam evaporation and patterned using a lift-off process. The total metal thickness is 750\,nm, optimized to ensure high surface conductivity due to the copper layer while minimizing electromagnetic wave penetration loss. The gold capping layer serves as a protective barrier against copper oxidation, while the intermediate chromium layers enhance adhesion between the metals and the underlying dielectric. Once patterning is complete, the RIS is carefully diced from the alumina substrate using a dicing saw. The ground plane is then formed by evaporating a uniform metal layer—identical to the top-layer stack—onto the opposite side of the substrate, with the edges protected using Kapton tape to prevent unintended metal deposition and leakage to the patterned top side. As the final step, a 250\,nm-thick silicon dioxide (SiO\textsubscript{2}) preservation layer is deposited via sputtering and patterned using a lift-off process. This layer is selectively removed from the DC pad areas located at the top and bottom edges of the RIS surface to allow electrical connectivity with external biasing wires. It is also removed from the VO\textsubscript{2} junctions to avoid any impact on switch operation. 

\subsection*{B) Measurement Setup}
\label{sec:switch_fab}

\begin{figure}[b]
    \centering
    \includegraphics[width=0.8\linewidth]{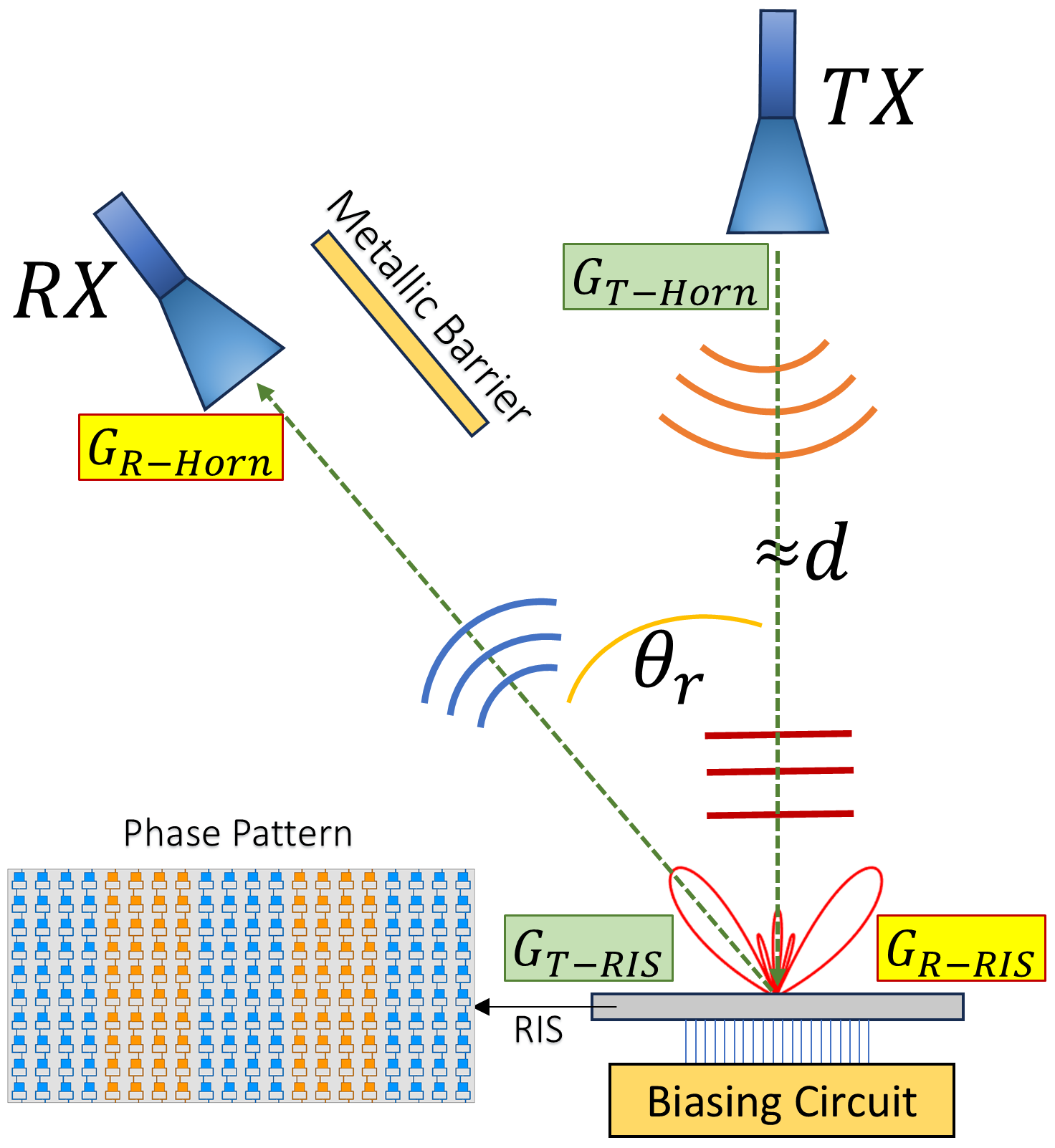} 
\caption{Schematic illustration of the measurement setup.}

    \label{fig:measurement_setup}
\end{figure}

Fig.~\ref{fig:measurement_setup} schematically illustrates the measurement setup. The RIS is placed 29~cm away from a transmitting horn antenna, which operates over the 26--40~GHz frequency band and provides an average gain of 12.5~dB. This separation ensures that both the RIS and the horn lie in each other's far-field, as the far-field condition is satisfied when the distance exceeds \(2D^2/\lambda\), where \(D\) is the aperture dimension of the larger element. At this distance, the incident wavefront impinging on the RIS is approximately planar, with only minor phase variations across its surface, as shown previously in Fig.~\ref{fig:unitcell_sparams_b}. The horn’s H-plane is aligned with the RIS steering plane to maintain linear polarization. The RIS reflection behavior is dynamically configured via a biasing circuit that modifies the local phase distribution to steer the reflected beam.
\begin{figure}[t]
    \centering
    \subfloat[]{%
        \includegraphics[width=0.8\linewidth]{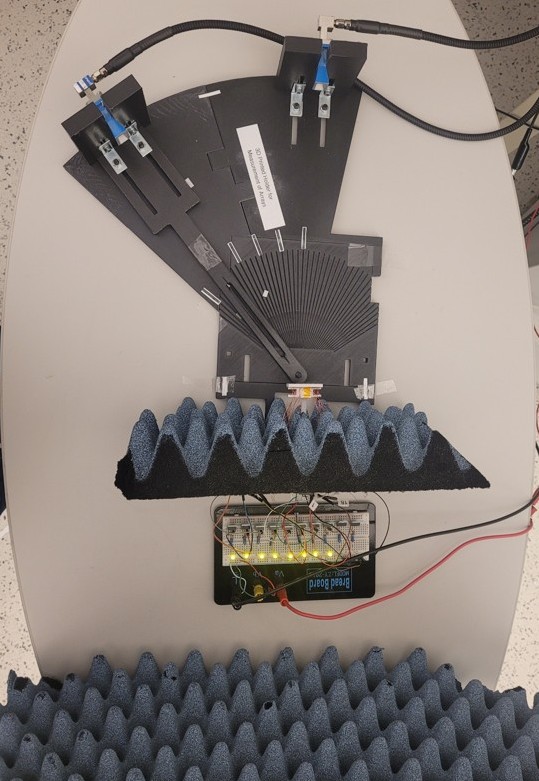}%
        \label{fig:meas_setup_photo}
    }\\[-0.1ex]  
    \subfloat[]{%
        \includegraphics[width=0.8\linewidth]{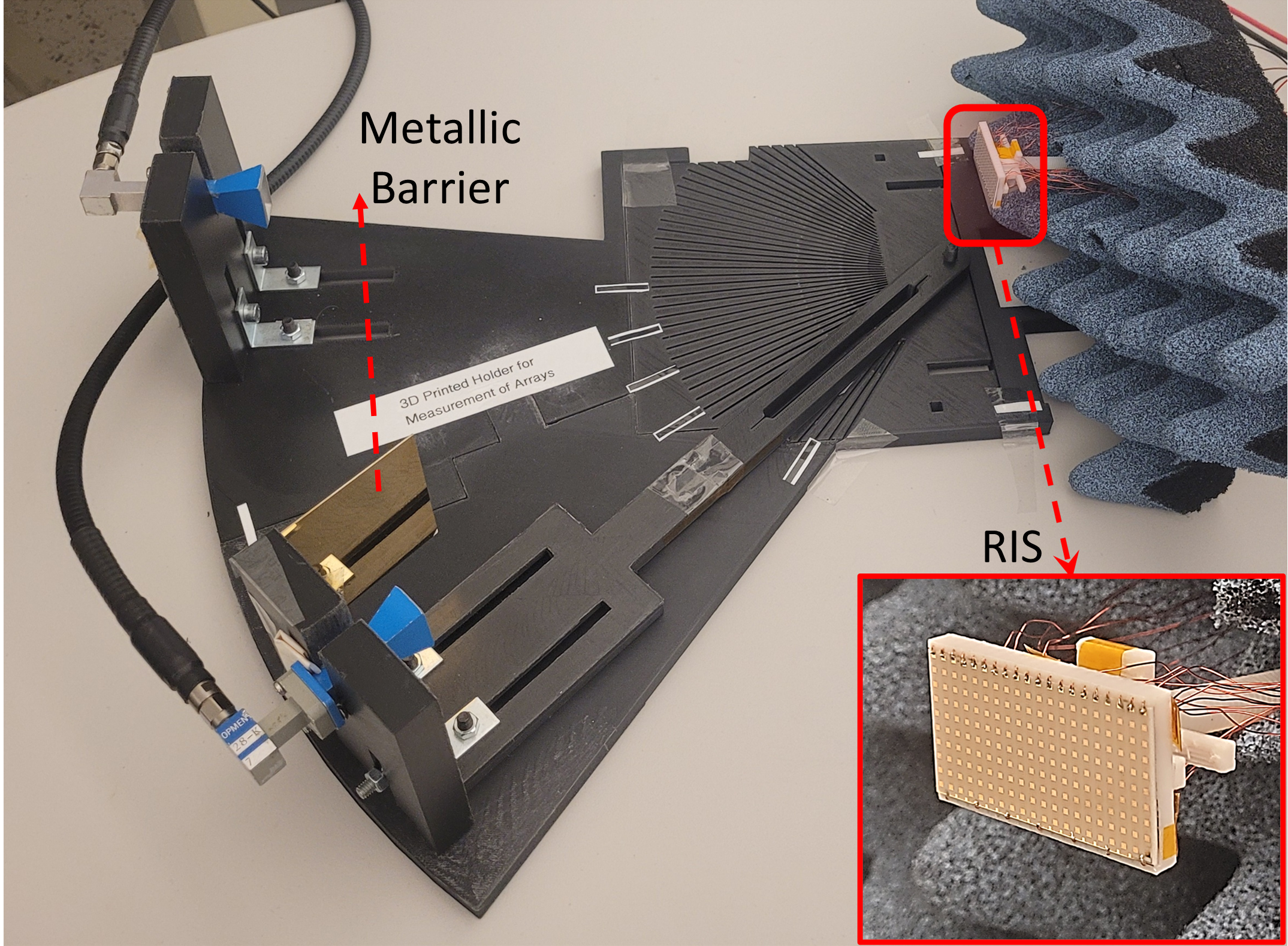}%
        \label{fig:meas_setup_schematic}
    }
   \caption{Measurement setup: (a) Top view; (b) Side view.}

    \label{fig:ris_measurement_stacked}
    \vspace{-0.5em}
\end{figure}

A second horn antenna, identical to the transmitting one, is positioned at an equal distance from the RIS and serves as the receiving element. Its angular position is adjustable to align with the direction of the reflected beam for each programmed RIS configuration. For each case, the receiving horn is rotated about the expected reflection angle to capture the direction of maximum received power. A metallic barrier is placed adjacent to the receiving antenna to suppress direct coupling between the two horns and to isolate the contribution of the RIS. The gain characteristics of the transmitting and receiving horns, as well as the RIS, are summarized in Fig.~\ref{fig:measurement_setup} and will be referenced later in Section~\ref{sec:measurement_results} to support the performance analysis. This setup facilitates the characterization of the RIS across the designated frequency band and enables precise identification of the angle corresponding to maximum signal reception. The subsequent section details the practical implementation of this configuration, including the biasing and measurement procedures for the microfabricated RIS.

Fig.~\ref{fig:ris_measurement_stacked} illustrates the actual measurement setup. A custom-designed, 3D-printed support structure fabricated from PLA (polylactic acid) includes two arms—one fixed for holding the transmitting horn antenna and the other rotatable for positioning the receiving horn antenna—as well as a platform for securely mounting the RIS. This structure ensures precise alignment between the antennas and the RIS. To mitigate background reflections and confine the antenna illumination exclusively to the RIS, an RF absorber is placed behind the setup, effectively minimizing undesired scattering from the rear side of the RIS during measurements. The biasing wires are routed behind the RIS ground plane, where they remain electromagnetically shielded from the incident wave to avoid RF interference. These wires pass through a dedicated opening in the absorber and exit from the rear of the setup, where they are connected to the external biasing circuit used for controlling the reflection phase distribution.

The three RIS patterns designed in this work require independent control of up to nine active columns. Consequently, the biasing system must be capable of selectively driving at least nine columns of unit cells within the RIS. To meet this requirement, a dedicated biasing circuit was developed. The circuit used for each column is identical to the one employed in the supercell biasing test, as described in Fig.~\ref{fig:madar}.

\textbf{Power consumption:} Each column of the array is driven at approximately 40~V and 75~mA. 
The intrinsic power consumption of the VO$_2$ switches is approximately 17~mW per element. 
Assuming that half of the 200 switches are in the ON state for a given coding pattern, this corresponds to a total power consumption of about 1.7~W for the full array.

\section{Measurement and Results}
\label{sec:measurement_results}

Based on the previously described measurement setup, the two coaxial ports of the vector network analyzer (VNA) were connected to the horn antennas. The $S_{21}$ parameter was measured for each designed phase pattern under two conditions: (i) with the RIS configured to impose the intended phase distribution, and (ii) with the RIS in an inactive state. These measurements are particularly informative, as $|S_{21}|^2$ represents the ratio of received power at the VNA's receiving port to the transmitted power at its source port. By evaluating $S_{21}$, the influence of the RIS on the received power at the targeted steering angle can be quantitatively assessed.

The expected power ratio can be estimated using the Friis transmission equation applied sequentially: first for the link between the transmitting horn and the RIS, and then for the link between the RIS and the receiving horn. This estimation assumes proper impedance matching between the horn antennas and the VNA coaxial ports across the entire frequency band.

\begin{equation}
\label{eq:power_ratio}
\begin{aligned}
\frac{P_R}{P_T} &=
G_{\text{T-horn}} \cdot G_{\text{R-RIS}}
\left( \frac{\lambda}{4\pi d} \right)^2 \\
&\quad \times
G_{\text{R-horn}} \cdot G_{\text{T-RIS}}
\left( \frac{\lambda}{4\pi d} \right)^2
\end{aligned}
\end{equation}

\noindent
In this expression, $G_{\text{T-horn}}$ and $G_{\text{R-horn}}$ denote the gains of the transmitting and receiving horn antennas, respectively, while $G_{\text{T-RIS}}$ and $G_{\text{R-RIS}}$ represent the effective transmitting and receiving gains of the RIS. The parameter $\lambda$ is the free-space wavelength, and $d$ is the distance between the horn antennas and the RIS, assumed to be equal for both the transmit and receive paths.

\begin{figure*}[t]
\centering
\vspace{-0.3cm}

\subfloat[]{
\includegraphics[width=0.3\textwidth]{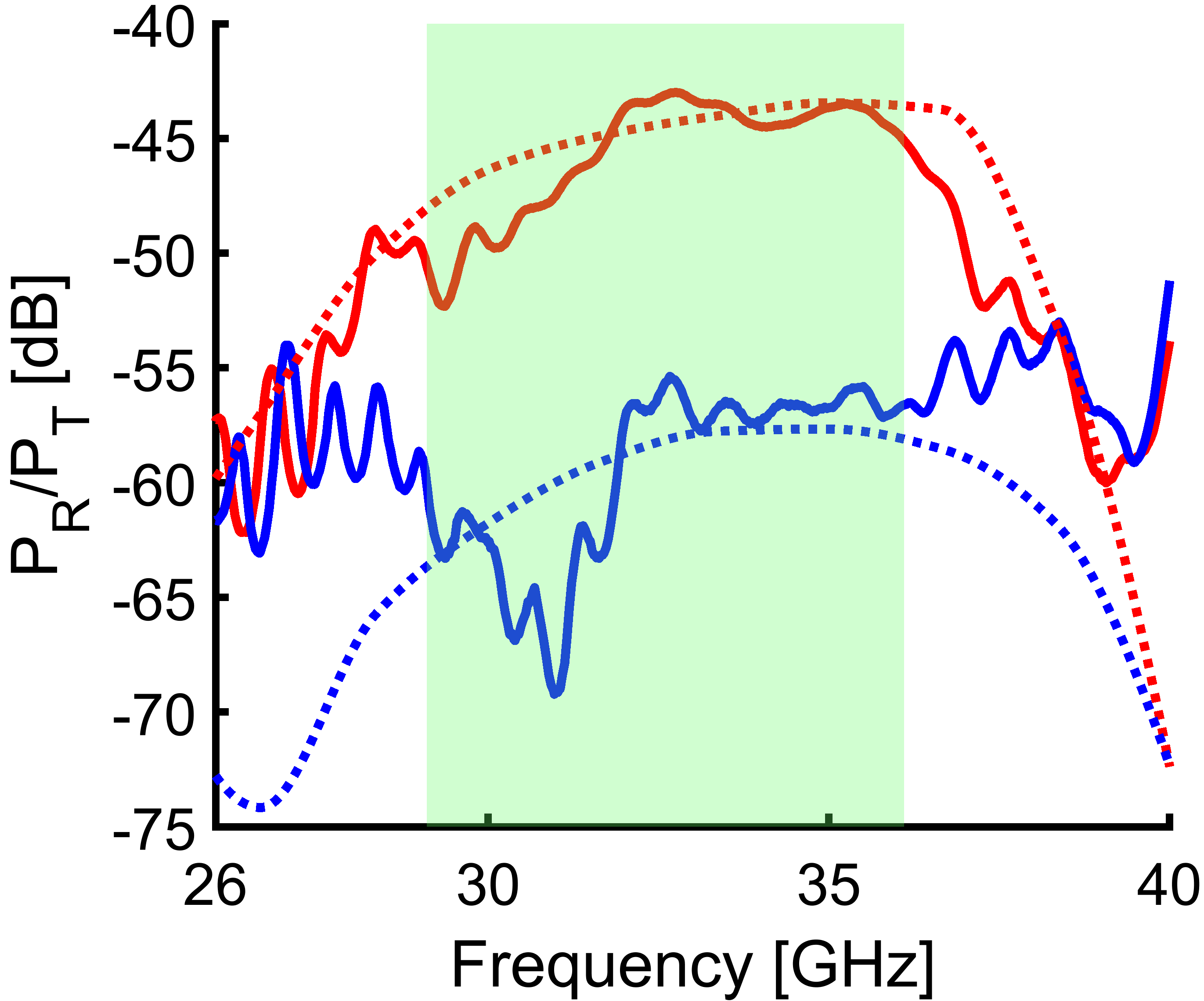}
\label{fig:top1}
}
\hfill
\subfloat[]{
\includegraphics[width=0.3\textwidth]{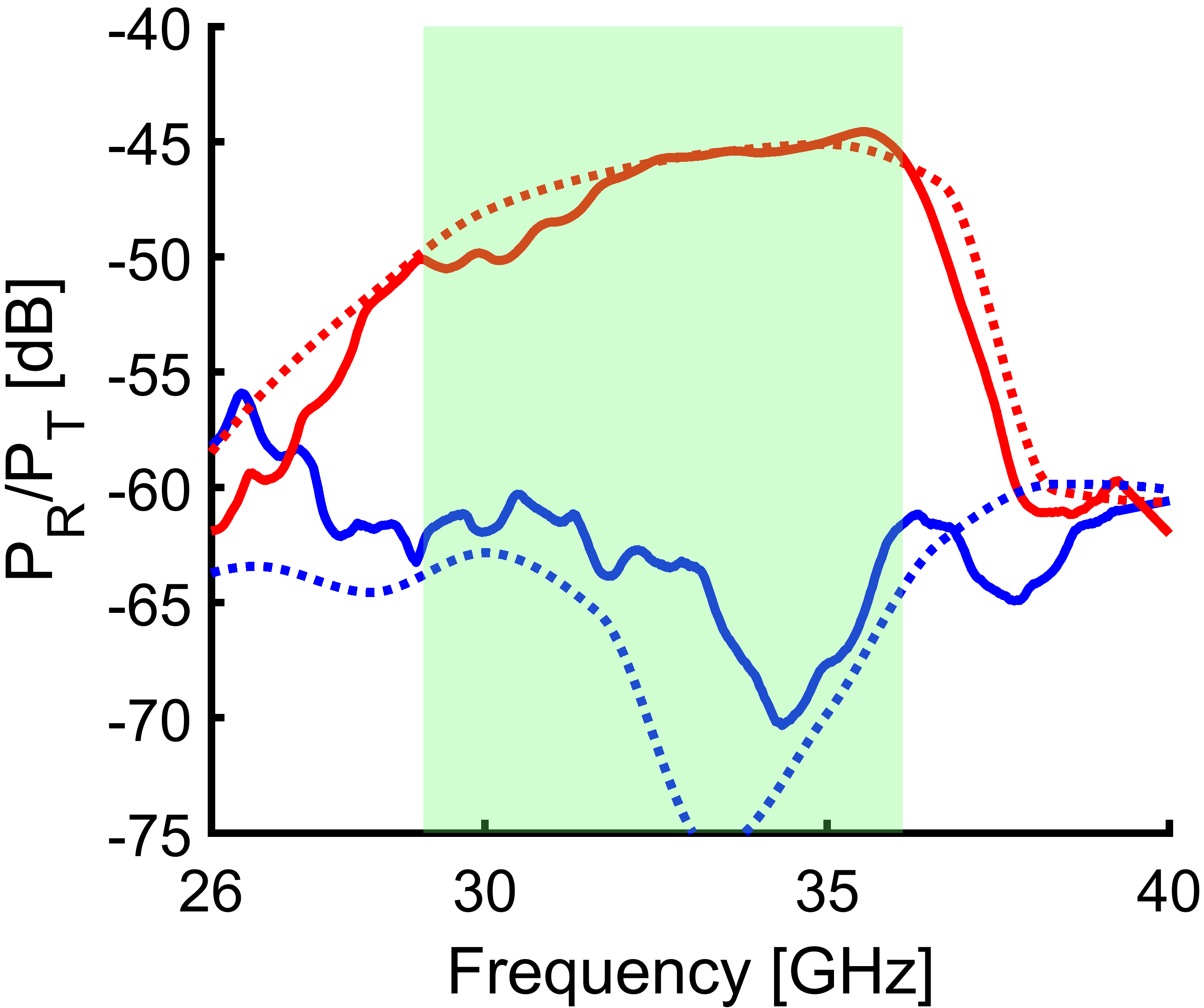}
\label{fig:top2}
}
\hfill
\subfloat[]{
\includegraphics[width=0.3\textwidth]{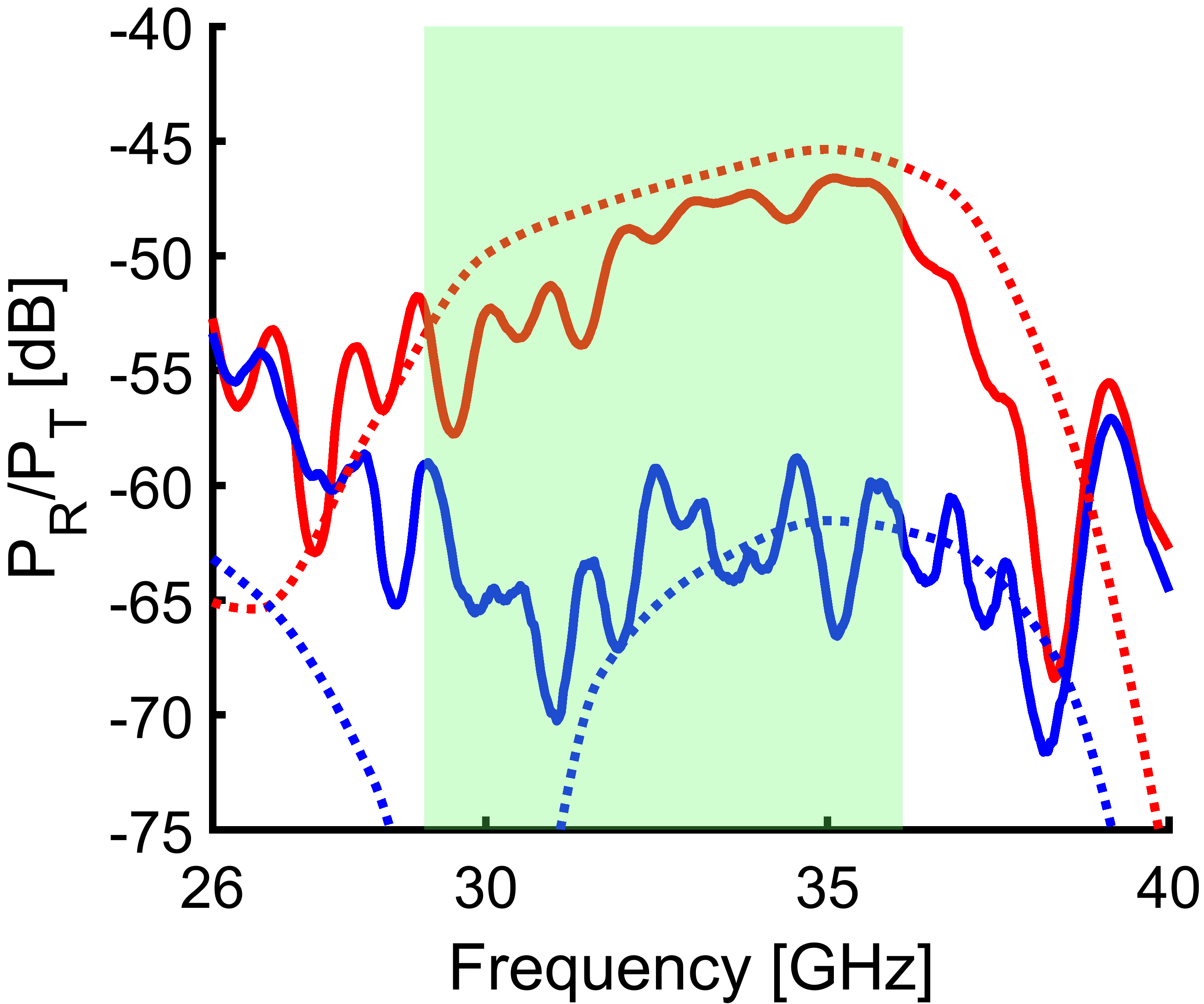}
\label{fig:top3}
}

\vspace{0.1cm}  

\subfloat[]{
\includegraphics[width=0.3\textwidth]{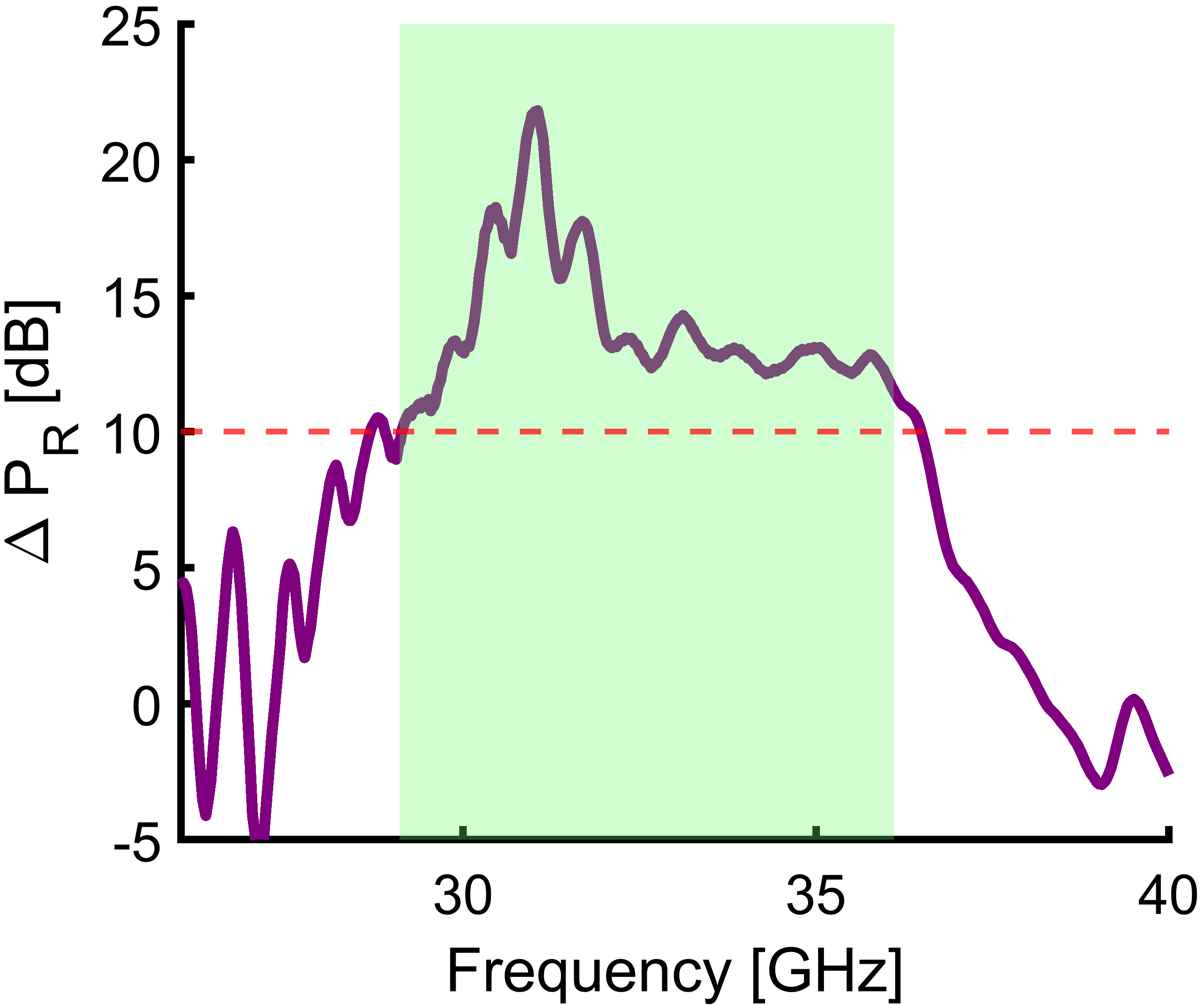}
\label{fig:bottom1}
}
\hfill
\subfloat[]{
\includegraphics[width=0.3\textwidth]{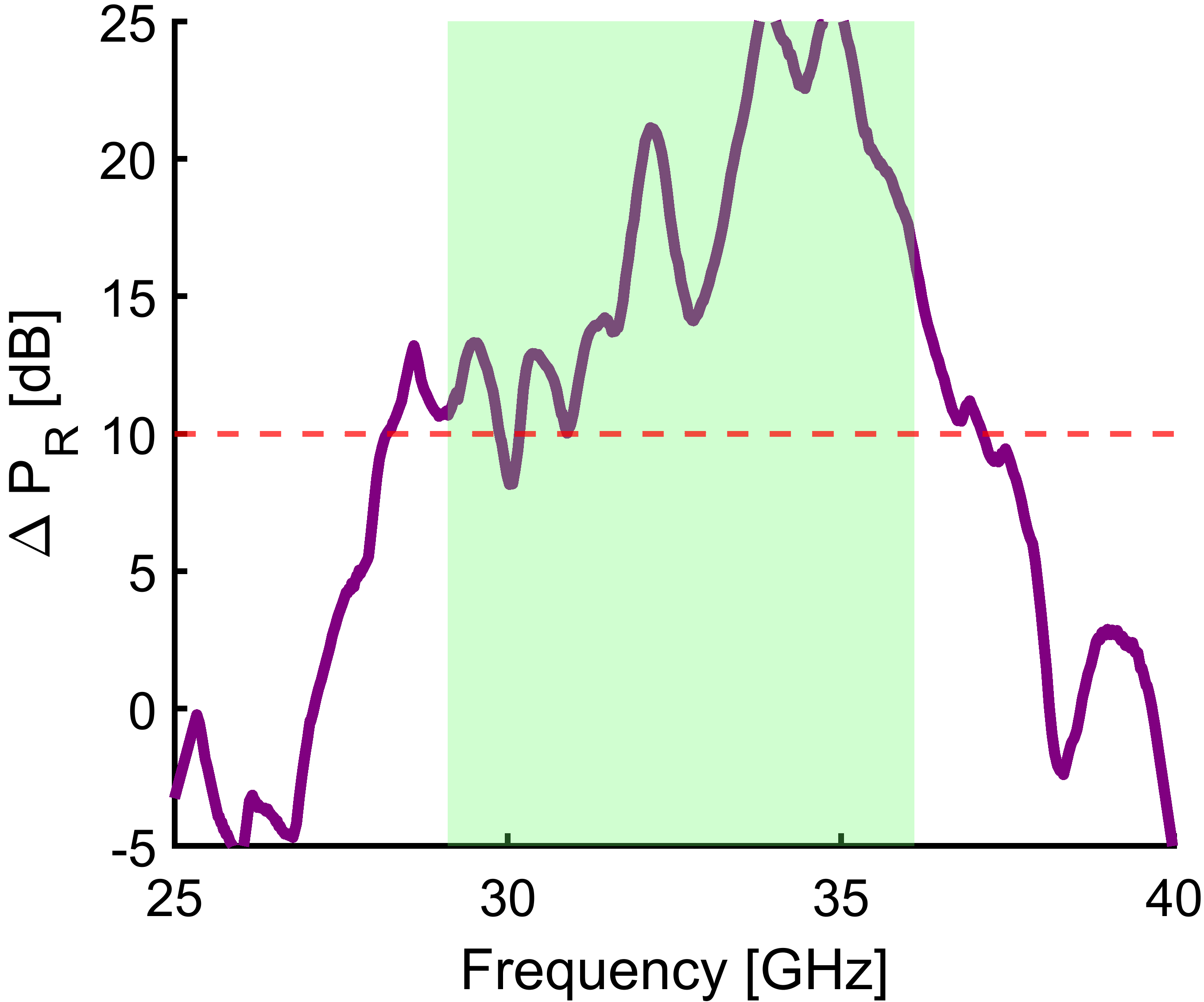}
\label{fig:bottom2}
}
\hfill
\subfloat[]{
\includegraphics[width=0.3\textwidth]{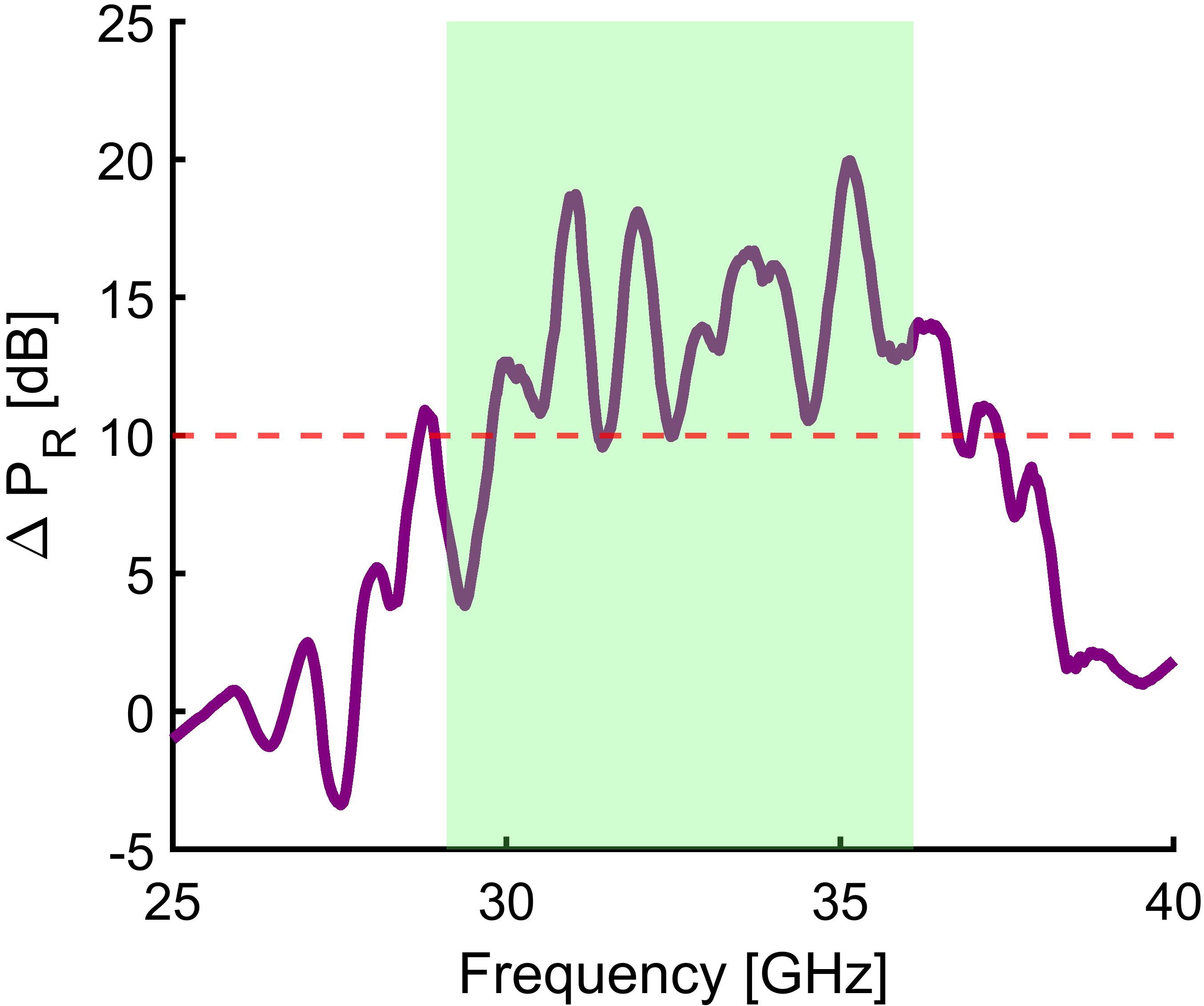}
\label{fig:bottom3}
}

\caption{Measured and semi-numerical results of the received-to-transmitted power ratio for RIS patterns designed to reflect beams at (a)~$40^\circ$, (b)~$50^\circ$, and (c)~$60^\circ$ across 26--40\,GHz, under both RIS ON and OFF conditions. (d)--(f) Directional gain enhancement for each RIS pattern. Legend: 
\protect\tikz[baseline=-0.6ex]\protect\draw[red, thick] (0,0)--(0.6,0); Meas. RIS ON, 
\protect\tikz[baseline=-0.6ex]\protect\draw[blue, thick] (0,0)--(0.6,0); Meas. RIS OFF, 
\protect\tikz[baseline=-0.6ex]\protect\draw[red, thick, dash pattern=on 3pt off 2pt] (0,0)--(0.6,0); Semi-num. RIS ON, 
\protect\tikz[baseline=-0.6ex]\protect\draw[blue, thick, dash pattern=on 1pt off 1pt] (0,0)--(0.6,0); Semi-num. RIS OFF.}

\label{fig:top-bottom-grid}
\vspace{-0.15cm}
\end{figure*}

The receiving gain of the RIS, $G_{\text{R-RIS}}$, is given by $G_{\text{R-RIS}} = \frac{4\pi}{\lambda^2} \cdot A_{\text{eff}}$, where $A_{\text{eff}} = \eta \cdot A_{\text{phys}}$. Here, $\eta$ is the aperture efficiency of the RIS during reception, which may vary with frequency, and $A_{\text{phys}}$ is its physical aperture area.

Fig.~\ref{fig:top-bottom-grid}\subref{fig:top1}--\subref{fig:top3} present the ratio of received to transmitted power at the angle corresponding to the maximum received power for three RIS patterns designed to steer the beam toward $40^\circ$, $50^\circ$, and $60^\circ$, respectively. Results are shown for both the activated and deactivated phase pattern conditions across the 26--40\,GHz frequency range.

To estimate the expected power ratio, a semi-numerical analysis was conducted based on \eqref{eq:power_ratio}, which computes the received-to-transmitted power ratio over the full frequency band. This hybrid approach combines full-wave and analytical methods to evaluate the system response. Specifically, the gain of the horn antennas was obtained through full-wave simulations using HFSS, based on their exact physical dimensions. The RIS transmitting electric far-field pattern was computed by combining full-wave simulation results of the unit cell with numerically implemented array synthesis, as described in Section~IV. This approach is referred to as the semi--full-wave method. The receiving gain of the RIS was computed analytically using the effective aperture formula. These components were then combined to evaluate the frequency-dependent received-to-transmitted power ratio.

As observed, the experimental power ratio under the activated RIS patterns shows good agreement with the numerically predicted values, with discrepancies becoming more pronounced toward the lower end of the predefined bandwidth. This deviation is likely due to the simplified assumption that the aperture efficiency factor $\eta$ is equal to unity. This assumption neglects the actual frequency-dependent efficiency of the RIS in receiving mode. At lower frequencies, the effective aperture efficiency is expected to decline as a result of increased fringing field effects, especially when the physical dimensions of the RIS become comparable to the operating wavelength. 

Furthermore, the semi–full-wave simulation used in this analysis is based on periodic boundary conditions for the unit cells. In practice, however, not all unit cells are surrounded by neighbors operating in the same state, and this approximation does not capture all inter-element coupling effects. Such coupling can influence RIS performance and reduce the effective operational bandwidth. Considering these factors, the observed discrepancies between the measured and calculated results are well justified.

When the RIS patterns are deactivated, the received power ratio is expected to be very low, as the RIS no longer reflects the signal toward the measurement direction. In this condition, the influence of cross-talk between the transmitting and receiving horn antennas, along with environmental noise, becomes more significant. Although the metallic barrier in the setup effectively reduces cross-talk, it does not completely eliminate it across the full frequency range. Consequently, the experimental measurements under the deactivated pattern condition show deviations from the semi-numerical values at certain frequencies, although the overall frequency-dependent trend remains consistent.

At each frequency point of measurement, all parameters in equation~\eqref{eq:power_ratio}—including the gains of the horn antennas, the wavelength~$\lambda$, and the distance~$d$—remain constant across both measurement conditions (RIS active and inactive). The only varying parameter is the directional transmitting gain of the RIS, $G_{\text{T-RIS}}$. Therefore, any variation in received power can be attributed solely to the influence of the RIS phase pattern. When activated, the RIS directs energy more efficiently toward the target direction, increasing $G_{\text{T-RIS}}$ and enhancing the received power. Consequently, the power difference between the activated and deactivated phase patterns corresponds to the RIS-induced gain enhancement, measured in dB. Fig.~\ref{fig:top-bottom-grid}\subref{fig:bottom1}--\subref{fig:bottom3} illustrate this difference. A gain improvement exceeding 10\,dB is observed across nearly the entire predefined frequency band—and even beyond—for all patterns, with higher values at certain frequencies. Collectively, the three phase configurations exhibit a gain enhancement ranging from 10 to 20\,dB over the 30.2--36.2\,GHz frequency range, corresponding to a relative bandwidth of more than 18\% centered around 33\,GHz.

 In general, the received power is higher for reflection angles of $40^\circ$ and $50^\circ$, as the unit cells were specifically designed to provide the required phase shifts for these directions. Nonetheless, the RIS still demonstrates substantial gain enhancement at $60^\circ$. These results confirm that the RIS effectively enhances the reflected beam gain in the intended directions, in alignment with the design objectives.

To mitigate the impact of measurement system noise and environmental interference—particularly significant when measuring low power ratios—the data were averaged using a 2\% moving frequency window and subsequently smoothed.

\begin{figure}[H]
    \centering
    \includegraphics[width=0.38\textwidth]{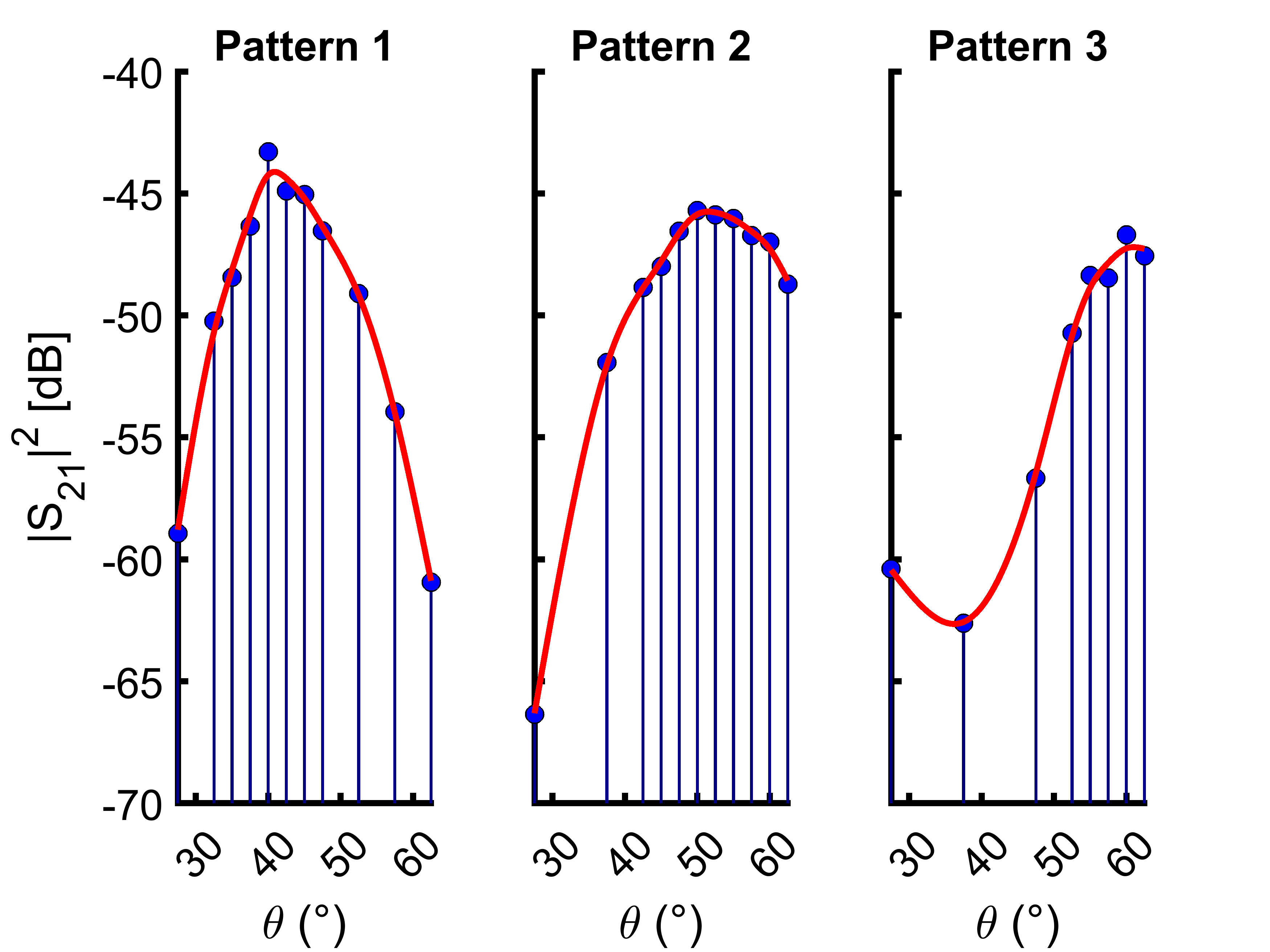}
   \caption{Measured received-to-transmitted power ratio around the designed reflection angles at the center frequency.}

    \label{fig:meas_power_ratio}
\end{figure}

\begin{figure}[H]
    \centering
    \includegraphics[width=0.38\textwidth]{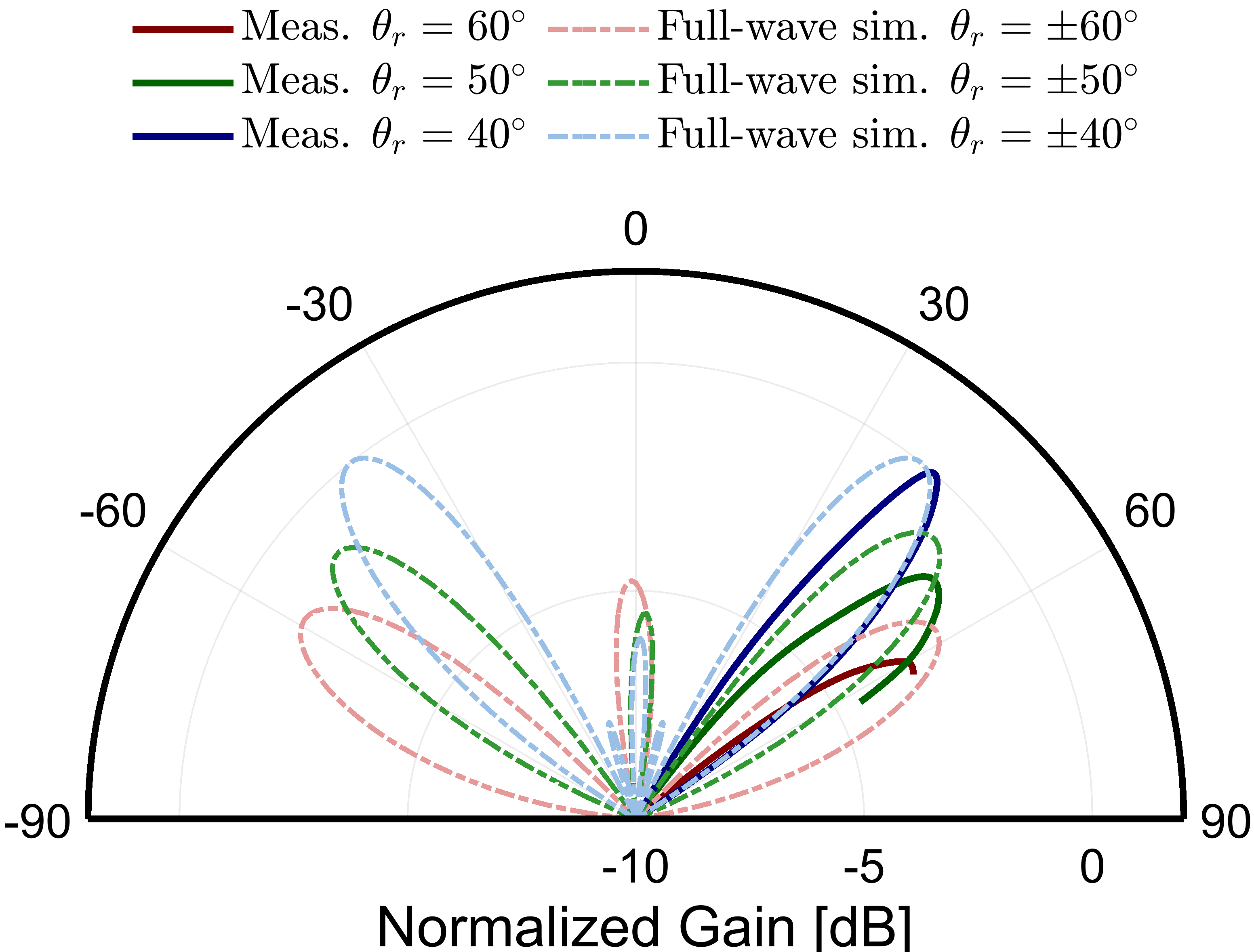}
    \caption{Measured and simulated normalized angular gain patterns for each RIS configuration.}
\vspace{-8pt}
    \label{fig:meas_angular_dist}
\end{figure}

The received-to-transmitted power ratio for each phase pattern was measured over a limited angular range (27.5$^\circ$ to 62.5$^\circ$) around the designed reflection directions, as shown in Fig.~\ref{fig:meas_power_ratio}. The results demonstrate that the RIS effectively maximizes reflected power toward the intended directions. These measurements were conducted at the center frequency; thus, variations in received power across different angles can be solely attributed to changes in the RIS's transmitting gain. Consequently, the normalized angular gain pattern of the RIS can be extracted from these measurements to highlight the angular response of each configuration. In Fig.~\ref{fig:meas_angular_dist}, the measured patterns are normalized to the maximum power ratio value and compared to their corresponding gain patterns extracted from full-wave simulations.

\section{Discussion}
\label{sec:discussion}

The proposed RIS employs an extremely miniaturized active PCM region that occupies only a negligible fraction of the unit-cell area, while still achieving strong and stable electromagnetic performance owing to the use of a high-quality microfabrication process. As a result of the drastically reduced tunable area, the design enables denser phase sampling through a smaller unit-cell size. In particular, although the present array aperture is smaller than that of the screen-printed implementation reported in~\cite{a71} and substantially smaller than the large-area screen-printed surface reported in~\cite{vo3}, it achieves higher gain enhancement and a wider beam-scanning range than the former, and comparable gain enhancement to the latter, despite the much larger physical size of the latter. These results indicate that the proposed design preserves wideband operation and does not sacrifice bandwidth despite aggressive miniaturization of the active region. Moreover, the column-wise electrical control enabled by the biasing network integrated into the unit-cell design allows programmable beam steering over multiple angles, which was not achievable in an earlier implemented RIS based on a microfabricated VO$_2$ switch that relied on global thermal actuation, as reported in~\cite{vo4}. Collectively, the results of this work demonstrate the potential of PCM switches when monolithically integrated into RIS architectures, particularly the combination of high-quality RF performance and highly miniaturized switch footprints enabled by microfabrication and careful unit-cell design.

While Table~\ref{tab:comparison} provides a performance-oriented comparison within the class of VO$_2$-based reconfigurable surfaces, it does not capture the fundamental integration and scalability constraints that differentiate VO$_2$ from other commonly used switching technologies in reconfigurable surfaces. These aspects are explicitly addressed in the integration-focused comparison presented in Table~\ref{tab:tech_compare}, which contrasts PIN diode-, MEMS-, and VO$_2$-based implementations in terms of switch footprint, unit-cell size, and integration approach.

As summarized in Table~\ref{tab:tech_compare}, PIN diode-based RIS implementations rely on discrete, solder-mounted components integrated within multilayer printed circuit boards. The relatively large physical footprint of commercial PIN diodes, together with the need for vias, bias routing, and hybrid assembly, limits achievable integration density and constrains unit-cell miniaturization at millimeter-wave frequencies. As a result, reported PIN-diode-based implementations typically operate with unit-cell dimensions on the order of $\lambda/2$ to $\lambda/3$ in the mmWave regime. At higher frequencies, the switch footprint becomes increasingly comparable to the wavelength, further complicating dense spatial sampling and highlighting the fundamental scalability limitations of this approach, as also implied in~\cite{cite1,cite2}.

\begin{table*}[t]
\centering
\caption{Comparison of VO$_2$-Based Reconfigurable Surface Implementations for Beam Steering}
\label{tab:comparison}
\footnotesize
\resizebox{\textwidth}{!}{%
\begin{tabular}{|c|c|c|c|c|c|c|c|c|c|c|}
\hline
\makecell{Work} & 
\makecell{Unit Cell \\ Size and \\ Array Layout} & 
\makecell{Tunable Area \\ Fraction} & 
\makecell{Unit Cell \\ Loss \\ @ Center Freq.} & 
\makecell{Maximum \\ Steering \\ Angle} & 
\makecell{Meas. Freq. \\ Band (GHz) \\ (FF Gain Enh.)} & 
\makecell{Far-Field Gain \\ Enhancement} & 
\makecell{Fabrication \\ Technology} & 
\makecell{Type} & 
\makecell{Actuation \\ Method} & 
\makecell{Phase \\ Change}
\\ \hline

\makecell{\cite{a70}} & 
\makecell{$\sim\lambda$/3.7 \\ (1.08$\lambda$$\times$1.08$\lambda$)} & 
\makecell{2.89\%} & 
-3 dB & 
$\pm$22° & 
100 & 
7.8 dB & 
Microfabrication & 
Transmittive & 
T-MIT / Column control & 
\makecell{Continuous \\ (59°)}
\\ \hline

\makecell{\cite{a71}} & 
\makecell{$\sim\lambda$/4.72 \\ (4.24$\lambda$$\times$4.24$\lambda$)}& 
\makecell{3.47\%} & 
-3.1 dB & 
$\pm$45° & 
\makecell{23.5 \\ (claimed over \\ 23.5--29.5~GHz)} & 
4.9 dB & 
Screen Printing & 
Reflective & 
E-MIT / Column control & 
\makecell{Discrete \\ ($\sim$180°)}
\\ \hline

\makecell{\cite{vo3}} & 
\makecell{$\sim\lambda$/1.78 \\ (28.6$\lambda$$\times$17.8$\lambda$)} & 
\makecell{1.7\%} & 
$\sim$-2 dB & 
$\sim\pm$70° & 
95--105 & 
\makecell{$ 10$ dB$\leq$} & 
Screen Printing & 
Reflective & 
E-MIT / Column control & 
\makecell{Discrete \\ ($\sim$180°)}
\\ \hline

\makecell{\cite{vo4}} & 
\makecell{$\sim\lambda$/4 \\ $\sim$10$\lambda$-diameter} & 
\makecell{3.3\%}& 
$\sim$-2 dB & 
$\pm$10° & 
35 & 
Not reported & 
Microfabrication & 
Reflective & 
Global heating & 
\makecell{Discrete \\ ($\sim$180°)}
\\ \hline

\makecell{\textbf{This Work}} & 
\makecell{$\sim\lambda$/5.24 \\ (1.9$\lambda$$\times$3.8$\lambda$)}& 
\makecell{ 0.004\%} & 
-1.5 dB & 
$\pm$60° & 
30--36 & 
\makecell{$ 10$ dB$\leq$  \\ $\leq 20$ dB} & 
Microfabrication & 
Reflective & 
E-MIT / Column control & 
\makecell{Discrete \\ ($\sim$180°)}
\\ \hline

\end{tabular}%
}
\end{table*}

\begin{table*}[t]
\centering
\caption{Quantitative and Qualitative Comparison of Switching Technologies for mmWave Beam-Steering Reconfigurable Surfaces (Integration-Focused)}
\label{tab:tech_compare}
\footnotesize
\resizebox{\textwidth}{!}{%
\begin{tabular}{|c|c|c|c|c|c|c|}
\hline
\makecell{Work} &
\makecell{Switch\\Technology} &
\makecell{Unit-Cell\\Size ($\lambda$)} &
\makecell{Switch\\Footprint ($\mu$m$^2$)} &
\makecell{Center\\Frequency (GHz)} &
\makecell{Commercial Part\\(Rated $f_\mathrm{max}$)} &
\makecell{Integration\\Method}
\\ \hline

\makecell{\cite{cite13}} &
PIN diode &
$\sim\lambda/2.5$ &
$\sim660\times640$ &
26 &
\makecell{MA4AGP907\\($\sim$50~GHz rated)} &
\makecell{Solder-mounted chip PIN diode\\on multilayer PCB with vias\\(hybrid, non-planar)}
\\ \hline

\makecell{\cite{a1}} &
PIN diode &
$\sim\lambda/3$ &
$\sim750\times380$ &
26 &
\makecell{MADP-000907-14020P\\($\sim$70~GHz rated)} &
\makecell{Solder-mounted chip PIN diodes\\on multilayer PCB with vias\\(hybrid, non-planar)}
\\ \hline

\makecell{\cite{cite1}} &
PIN diode &
$\sim\lambda/2.5$ &
$\sim660\times340$ &
40 &
\makecell{MA4AGP907\\($\sim$50~GHz rated)} &
\makecell{Solder-mounted chip PIN diodes\\on multilayer PCB with vias\\(hybrid, non-planar)}
\\ \hline

\makecell{\cite{cite2}} &
PIN diode &
$\sim\lambda/2$ &
$\sim660\times340$ &
73 &
\makecell{MA4AGFCP910\\($\sim$50~GHz rated)} &
\makecell{Diagonally soldered chip PIN diodes\\on multilayer PCB with vias\\(hybrid, non-planar)}
\\ \hline

\makecell{\cite{mems1}} &
RF MEMS &
$\sim\lambda/2$ &
$\sim400\times150$ &
26.5 &
\makecell{N/A\\(custom RF MEMS switch)} &
\makecell{Monolithic multilayer\\surface-micromachined MEMS\\(wafer-level fabrication)}
\\ \hline

\makecell{\cite{mems2}} &
RF MEMS &
$\sim\lambda/2$ &
$\sim410\times90$ &
76.5 &
\makecell{N/A\\(custom RF MEMS switch)} &
\makecell{Hybrid MEMS chips mounted in PCB cavities\\with wire-bonded biasing\\(non-planar)}
\\ \hline

\makecell{\textbf{This Work}} &
VO$_2$ (E-MIT) &
\textbf{$\lambda/5.24$} &
\textbf{$\sim4\times20$} &
33 &
\makecell{N/A\\(custom VO$_2$ switch)} &
\makecell{Monolithic single-substrate thin-film integration\\(planar, density-scalable)}
\\ \hline

\end{tabular}%
}
\end{table*}

RF MEMS-based reconfigurable surfaces mitigate some of the loss and linearity limitations associated with semiconductor switches, but introduce distinct integration challenges. The fully monolithic MEMS implementation reported in~\cite{mems1} requires complex surface-micromachining processes, wafer bonding, and high-yield fabrication over large apertures, whereas the alternative MEMS-based architecture demonstrated in~\cite{mems2} relies on chip-level assembly, cavity mounting, and wire-bonded interconnects. As can be seen, most reported MEMS-based reconfigurable surfaces employ unit-cell dimensions on the order of $\lambda/2$, placing them closer to the reconfigurable reflectarray regime rather than a densely sampled RIS architecture.

In contrast, the proposed VO$_2$-based RIS exploits micron-scale, lithographically defined switches that are directly patterned within the unit-cell metallization. As a result, the active PCM region occupies only a negligible fraction of the unit-cell area—orders of magnitude smaller than those used in diode- and MEMS-based approaches as well as previously reported VO$_2$ implementations—allowing the unit cell to remain deeply subwavelength and providing ample margin for further miniaturization and scaling to higher frequencies, where the primary constraint shifts from switch footprint to the intrinsic high-frequency performance of the switching element. This monolithic, single-substrate thin-film integration minimizes parasitic effects, eliminates hybrid assembly steps, and provides a viable path toward scalable operation at higher millimeter-wave and sub-terahertz frequencies, where integration density and electromagnetic performance become the dominant considerations.

Moreover, despite the extremely small dimensions of the PCM switches, all switches in the fabricated array were functional, demonstrating the reliability required for large-scale, microfabrication-enabled RIS platforms. In the present column-addressed architecture, the failure of a single switch would disable its corresponding column, reducing the effective aperture by approximately 5\%. Future implementations can mitigate this effect through unit-cell-level access architectures, such as that introduced in our earlier work~\cite{a72}, to further enhance fault tolerance. Given the demonstrated process uniformity and fabrication precision, such defects are expected to remain rare as the array dimensions scale.

Finally, the operational robustness of VO$_2$-based switches is supported by the intrinsic sharpness of the metal--insulator transition near 68$^\circ$C. Practical packaging techniques can stabilize device temperature, and material engineering approaches can be employed to shift the transition temperature upward, enabling reliable operation under realistic environmental conditions~\cite{a75,a76}. Together, these considerations confirm the feasibility of robust, large-scale RIS implementations based on monolithically integrated VO$_2$ switches.

\section{Conclusion}
\label{sec:conclusion}

This paper presented the design, fabrication, and experimental validation of a monolithically integrated reconfigurable intelligent surface (RIS) based on electrically actuated VO$_2$ switches for millimeter-wave applications. Building on a previously proposed subwavelength unit cell~\cite{a73}, this work analyzed the switching-induced phase modulation through a surface-impedance framework and experimentally validated the phase and amplitude responses using integrated VO$_2$ switches in a WR-28 waveguide measurement setup. The measured electromagnetic behavior closely matched earlier results obtained using idealized open- and short-circuit switching conditions, thereby confirming the effectiveness of the VO$_2$ switching approach and its minimal impact on the unit-cell response.

Using the validated unit-cell design, a $10\times20$ RIS array was fabricated through a multilayer microfabrication process. Far-field measurements demonstrated programmable beam steering toward $40^\circ$, $50^\circ$, and $60^\circ$, with 10--20~dB of gain enhancement over an 18\% fractional bandwidth centered at 33~GHz. The measured radiation patterns were in good agreement with predictions from semi-numerical channel modeling. The column-wise electrical control enabled by serially biased VO$_2$ switches allowed dynamic reconfiguration of the spatial phase profile across the surface using a compact and integration-friendly biasing architecture.

Beyond these experimental demonstrations, this work establishes a fully monolithic, microfabricated RIS integrating 200 sub-4~$\mu$m VO$_2$ switches with consistent and reliable operation across the array. By leveraging lithographically defined, micron-scale PCM switches that occupy only a negligible fraction of the unit-cell area, the proposed architecture achieves deep subwavelength sampling, wideband phase contrast, and high-gain beam steering without the parasitic penalties and scalability limitations associated with hybrid diode-based or mechanically complex MEMS-based implementations. These results position microfabricated VO$_2$ switching as a compelling and scalable platform for future RIS architectures operating at higher millimeter-wave and sub-terahertz frequencies, where integration density, manufacturability, and electromagnetic performance become increasingly critical.

\bibliographystyle{IEEEtran}
\bibliography{ref2}

\end{document}